\pdfoutput=1
\documentclass[cits]{JINST}

\usepackage{array}
\usepackage{amsmath}

\title{Characterization and modeling of crosstalk and afterpulsing in Hamamatsu silicon photomultipliers}

\author{J. Rosado\thanks{Corresponding author.}\\
Departamento de F\'{i}sica At\'{o}mica, Molecular y Nuclear, Facultad de Ciencias F\'{i}sicas,\\
Universidad Complutense de Madrid, E-28040 Madrid, Spain\\
E-mail: \email{jaime\_ros@fis.ucm.es}}
\author{S. Hidalgo\\
Centro Nacional de Microelectr\'{o}nica, IMB-CNM (CSIC),\\
Campus UAB, E-08193 Bellaterra, Spain\\
E-mail: \email{salvador.hidalgo@csic.es}}

\abstract{%
The crosstalk and afterpulsing in Hamamatsu silicon photomultipliers, called Multi-Pixel Photon Counters (MPPCs), have been studied in depth.
Several components of the correlated noise have been identified according to their different possible causes and their effects on the signal.
In particular, we have distinguished between prompt and delayed crosstalk as well as between trap-assisted and hole-induced afterpulsing.

The prompt crosstalk has been characterized through the pulse amplitude spectrum measured at dark conditions.
The newest MPPC series, which incorporate isolating trenches between pixels, exhibit a very low prompt crosstalk,
but a small component remains likely due to secondary photons reflected on the top surface of the device
and photon-generated minority carriers diffusing in the silicon substrate.

We present a meticulous procedure to characterize the afterpulsing and delayed crosstalk through the amplitude
and delay time distributions of secondary pulses.
Our results indicate that both noise components are due to minority carriers diffusing in the substrate
and that this effect is drastically reduced in the new MPPC series as a consequence of an increase of one order of magnitude in the doping density of the substrate.

Finally, we have developed a Monte Carlo simulation to study the different components of the afterpulsing and crosstalk.
The simulation results support our interpretation of the experimental data.
They also demonstrate that trenches longer than those employed in the Hamamatsu MPPCs would reduce the crosstalk to a much greater extent.%
}

\keywords{%
Avalanche-induced secondary effects;
Photon detectors for UV, visible and IR photons (solid-state) (PIN diodes, APDs, Si-PMTs, G-APDs, CCDs, EBCCDs, EMCCDs, etc);
Detector modelling and simulations II (electric fields, charge transport, multiplication and induction, pulse formation, electron emission, etc)%
}

\begin{document}

\section{Introduction}\label{sec:intro}

Recent progresses in silicon photomultipliers (SiPMs) have allowed a notable reduction of the correlated noise:
crosstalk and afterpulsing. However these effects are still major drawbacks of SiPMs. Their characterization and
modeling are necessary for an appropriate processing and interpretation of data as well as for the development of
detectors with a further noise reduction.

Afterpulses originate from parasitic avalanches in a pixel that was previously fired, e.g., by an incident photon. They
are commonly attributed to the delayed release of carriers that were trapped in deep levels in the depletion layer
during the previous avalanche~\cite{Cova}. There may be various afterpulse components due to traps with different
lifetimes. Both the trigger probability and the charge multiplication of these secondary avalanches are lowered while
the pixel voltage is recovering from the last avalanche breakdown. Afterpulses distort the single photon counting and
increase the integrated charge registered for an input light pulse in a time-dependent way.

\begin{figure}[!t]
\centering
\includegraphics[width=.8\textwidth]{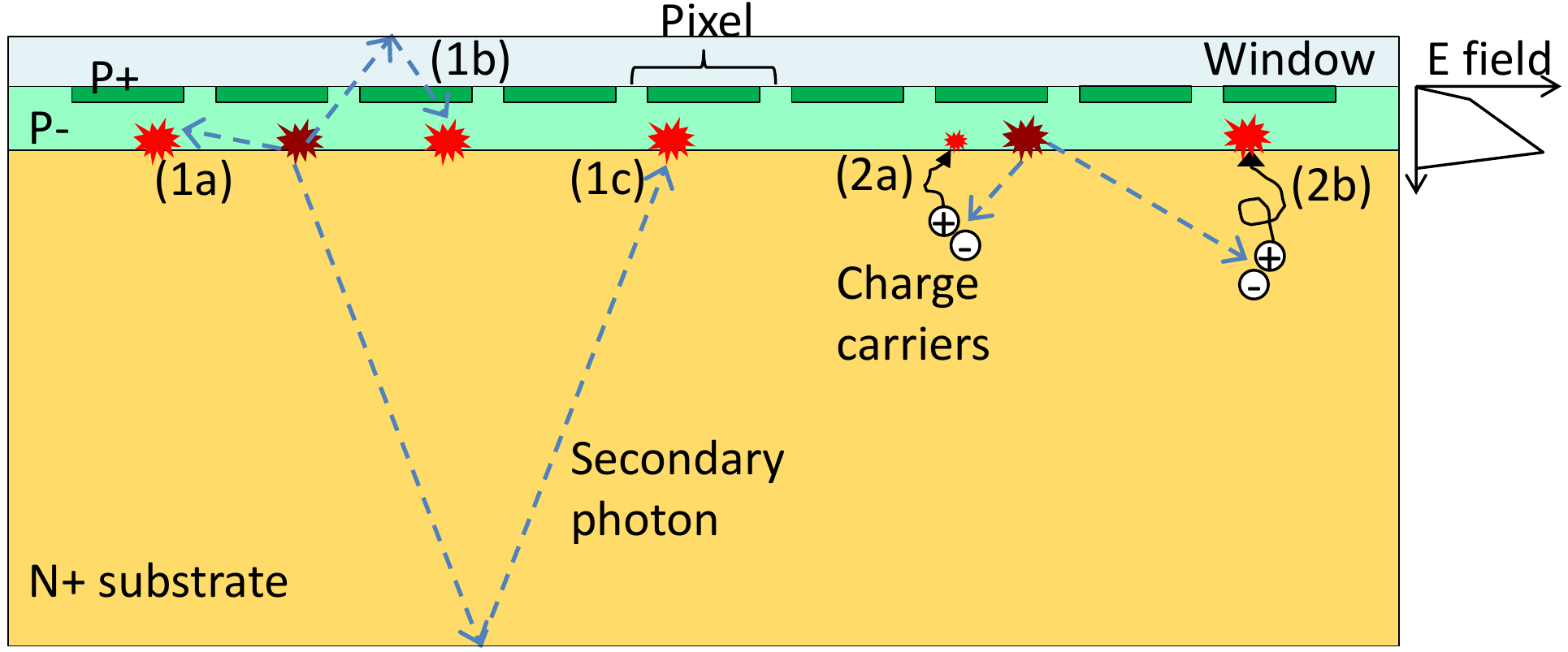}
\caption{Physical processes that cause correlated noise.
Primary avalanches are represented in dark red and secondary avalanches in bright red.
Optical crosstalk is due to secondary photons reaching
directly a neighboring pixel (1a) or after a reflection on either the top surface (1b) or the bottom surface (1c)
of the device. A minority carrier generated in the substrate may diffuse to the primary pixel (2a) or to a
neighboring one (2b) and produce an afterpulse or delayed crosstalk, respectively.
Trap-assisted afterpulsing (see text) is not illustrated in the figure.
The electric field intensity as a function of depth is shown on the right.} \label{fig:processes}
\end{figure}

The crosstalk is the correlated generation of avalanches in pixels different to the one where a primary avalanche took
place. Secondary photons emitted in the primary avalanche (see~\cite{Mirzoyan} and references therein) can reach
neighboring pixels, either directly or after a reflection, and trigger avalanches there almost simultaneously to the
primary one (figure~\ref{fig:processes}). Hence, this effect is called optical crosstalk. Secondary avalanches also
emit photons that can trigger more simultaneous avalanches in other pixels, giving rise to high-amplitude pulses even
at dark conditions. The crosstalk degrades the excellent photon counting resolution of SiPMs, unless the crosstalk
multiplicity is known and properly accounted for~\cite{Gallego}.

Besides, secondary photons absorbed in non-depleted regions of the device, especially the silicon substrate, produce
carriers that can diffuse to the depleted region of a pixel and trigger a secondary avalanche~\cite{Buzhan}. In the
case of the p on n structure shown in figure~\ref{fig:processes}, only holes originated from the n-type substrate are
accelerated upwards by the electrical field in the junction and able to trigger an avalanche. The delay time of this
secondary avalanche is dominated by the carrier diffusion time, which relies on the depth at which the secondary photon
was absorbed and the doping density of the substrate. Depending on whether the carrier reaches the primary pixel or any
other one, the event is classified as afterpulse or crosstalk, respectively. These two components are undistinguishable
at long delay time, but afterpulses with delay times shorter than the pixel recovery time have lower amplitude. On the
other hand, since crosstalk avalanches are not affected by the primary pixel recovery, they may have arbitrarily short
delay times. If the delay time is even shorter than the time resolution of the detection system, the output signal will
be basically the same as if the secondary photon had been absorbed within the depletion region of the pixel. According
to that, crosstalk events are classified as prompt or delayed.

In this work, we focused on the SiPMs fabricated by Hamamatsu Photonics K.K.~\cite{Hamamatsu}. This company named its
product Multi-Pixel Photon Counter (MPPC), which is a trademark. Hamamatsu has released several series of MPPCs based
on the p on n ``reverse junction'' structure of figure~\ref{fig:processes}, which was designed to optimize the photon
detection efficiency around 400~nm~\cite{Yamamoto}. The S10362-11 and S10362-33 series (not marketed now) are MPPCs
with active areas of 1~mm~$\times$~1~mm and 3~mm~$\times$~3~mm, respectively. They were fabricated with pixel pitches
of 25, 50 or 100~$\mu$m. The S12571 and S12572 series were produced with the same active areas as the former S10362-11
and S10362-33 series, but in addition to the 25, 50 and 100~$\mu$m pixel pitch types, MPPCs with pixel pitches of 10
and 15~$\mu$m were produced, offering larger dynamic range and shorter recovery time. A high fill factor (i.e., the
ratio of the active area of a pixel to its total area) ranging from 33\% for the 10~$\mu$m pixel pitch type to 78\% for
the 100~$\mu$m pixel pitch type was achieved. The most prominent progress with respect to the older series is a drastic
reduction of afterpulsing, which was attained thanks to the use of improved materials and wafer process technologies
according to Hamamatsu's reports. Recently, it has been released the S13360 series incorporating optical barriers,
called trenches, between pixels that drastically reduce crosstalk too. Low crosstalk and afterpulse MPPCs with active
areas of 1.3~mm~$\times$~1.3~mm, 3~mm~$\times$~3~mm and 6~mm~$\times$~6~mm, and with pixel pitches of 25 and 50~$\mu$m
are available. We characterized experimentally and modeled the afterpulsing and crosstalk for these three series of
MPPCs, studying the different noise components according to their physical causes. Solutions for a further reduction of
these effects are proposed.

The paper is structured as follows. The experimental method is described in section~\ref{sec:method}. The results on
prompt crosstalk are given and compared with model predictions in section~\ref{sec:CT}. The afterpulsing and delayed
crosstalk are characterized and modeled in section~\ref{sec:aft}. The possible causes of correlated noise and their
solutions are analyzed by means of Monte Carlo simulations in section~\ref{sec:MC}. The conclusions are in
section~\ref{sec:conclusions}.

\section{Experimental method}\label{sec:method}

In this work, all measurements were carried out at dark conditions and room temperature; the detector temperature
ranging from 25 to 30$^\circ$C. Pulses produced by thermal effect, called dark counts, were registered at a low rate of
$\lesssim1$~MHz. This allowed a characterization of the correlated noise on a single primary avalanche basis and almost
free from contamination due to coincident uncorrelated dark counts. Our technique is based on a waveform analysis to
identify individual pulses through their leading edges. We focused on the MPPCs with the smallest active area (i.e.,
1~mm~$\times$~1~mm and 1.3~mm~$\times$~1.3~mm), because they produce narrower and higher pulses, although some
3~mm~$\times$~3~mm active area MPPCs were also characterized.

We employed a Hamamatsu C12332 driver circuit for MPPCs~\cite{Hamamatsu}. It included a regulable power supply, a
high-speed operational amplifier and a temperature sensor. Output signals were registered by a digital oscilloscope
(Tektronix TDS5032B) and stored in binary files for later analysis. The oscilloscope was programmed to trigger randomly
1000 times per run with a time window of 3.2~ms and a sampling rate of 2.5~GS/s. A run contained typically around
$10^6$ pulses. Temperature variations during a run were less than $0.5^\circ$C and were neglected in the analysis, but
run-to-run temperature variations were corrected using the temperature coefficient of the breakdown voltage provided by
Hamamatsu.

The data analysis procedure was described in detail in~\cite{Gallego} and only a few upgrades were made since then. A
C++ code was developed to read the files generated by the oscilloscope's software and perform a waveform analysis. It
deconvolves the signal assuming an exponential response function with a time constant depending on the MPPC type. The
detectors of the S13360 series were found to produce pulses with a spike component of a few ns followed by a slow
component of tens of ns (see, e.g.,~\cite{Otono}). The deconvolution only removes the slow component in this case. A
pulse is identified when the deconvolved signal surpasses a discrimination threshold that is tuned for each MPPC and
bias voltage. In this way, pulses are separated with a resolution better than 10~ns. Furthermore, for every identified
pulse, an algorithm searches for double-peak structures in the signal over the threshold revealing very close pulses
that cannot be separated. The resolution to identify these piled-up pulses is estimated to be 3~ns, which determines
the minimum delay time to distinguish delayed from prompt crosstalk.

The pulse amplitude and time are measured at the maximum of the deconvolved pulse. In addition, the code calculates the
pulse height in the original signal after subtracting the baseline by means of an exponential extrapolation of the
signal in a small region prior to the pulse edge. The original pulse height has a lower statistical uncertainty because
it is less affected by electronic noise than the deconvolved pulse maximum, but the former cannot be applied to a pulse
that is closer than 20~ns to the preceding one.

We studied the dependence of the correlated noise on the MPPC gain, which is proportional to the overvoltage, i.e., the
difference between the bias voltage and the breakdown voltage. To do that, the breakdown voltage of each MPPC was
determined by zero extrapolation of the separation between the first and second peaks in the pulse amplitude spectrum
versus bias voltage~\cite{Gallego}. Both the deconvolved pulse amplitude and the original pulse height gave the same
breakdown voltage within uncertainties. The proportionality constant that relates overvoltage and gain was provided by
Hamamatsu.

\section{Characterization of prompt crosstalk}\label{sec:CT}

\subsection{Pulse amplitude spectra}\label{ssec:spectra}

We characterized the prompt crosstalk from either the deconvolved pulse amplitude spectrum or the pulse height
spectrum, although the latter had a better resolution in most cases. Only pulses that were separated from the previous
pulse by more than 120~ns were selected to calculate the spectrum. This prevented including pulses originated from
pixels that were still recovering, as they have lower gain and therefore lower probability to induce crosstalk. Cuts
were also applied to remove piled-up pulses and false pulses due to electronic noise exceeding the discrimination
threshold.

The relative areas of the peaks in the amplitude spectrum give the probability distribution of the prompt crosstalk
multiplicity, where the k-th peak corresponds to a multiplicity k-1. The overall crosstalk probability $\varepsilon$ is
thus defined by one minus the relative area of the first peak. More precisely, a small correction was made to account
for coincidences of uncorrelated events in a time interval of 3~ns ($\lesssim0.001$ probability), as was described
in~\cite{Gallego}.

\begin{figure}[t]
\centering
\includegraphics[width=.6\textwidth]{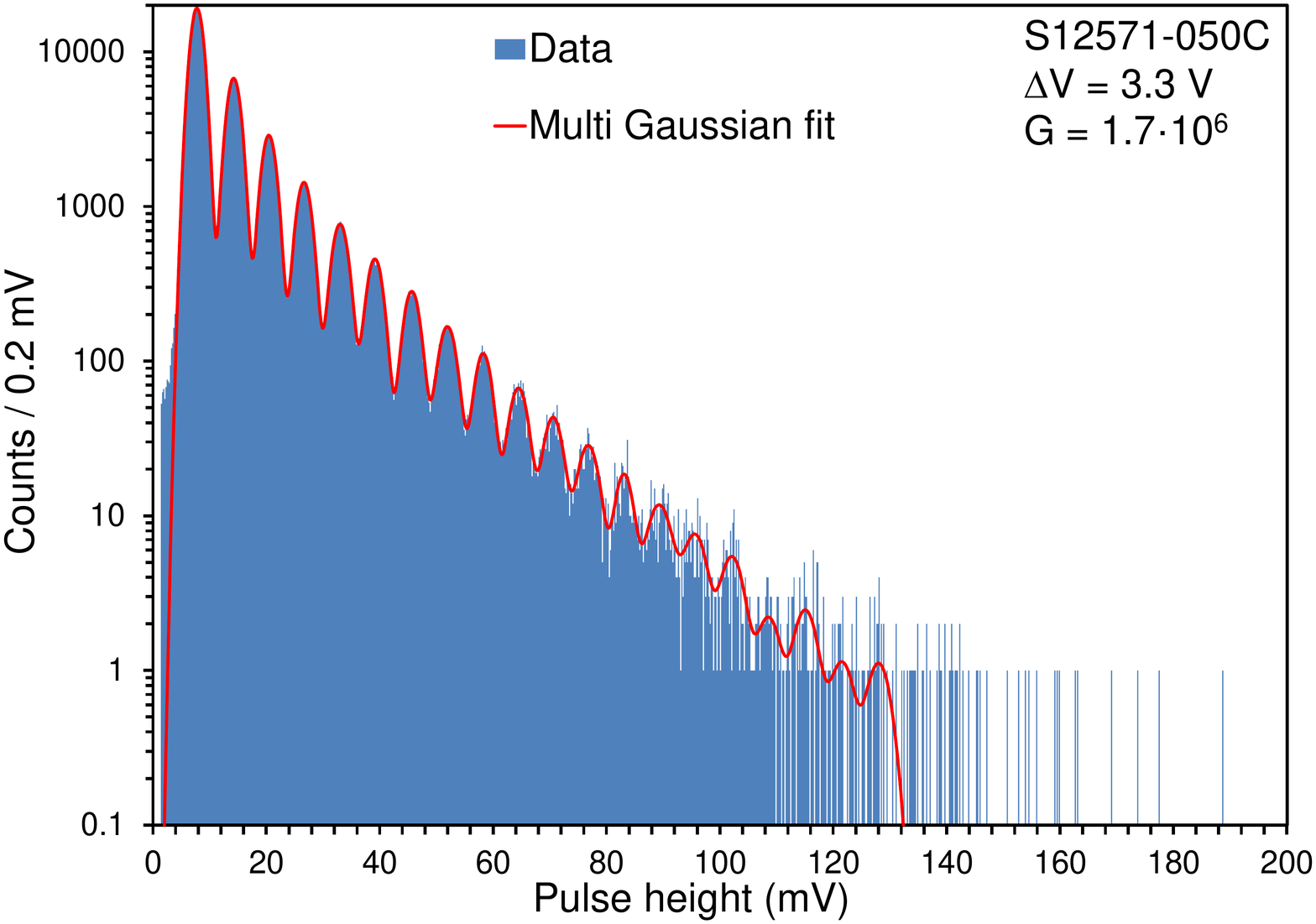}
\caption{Pulse height spectrum obtained for the S12571-050C MPPC at dark conditions.
Both a high probability and multiplicity of prompt crosstalk are observed.
Our waveform analysis allowed us to resolve at least 16 Gaussian-like peaks.}
\label{fig:spectrum_height_12571}
\end{figure}

\begin{figure}[t]
\centering
\includegraphics[width=.45\textwidth]{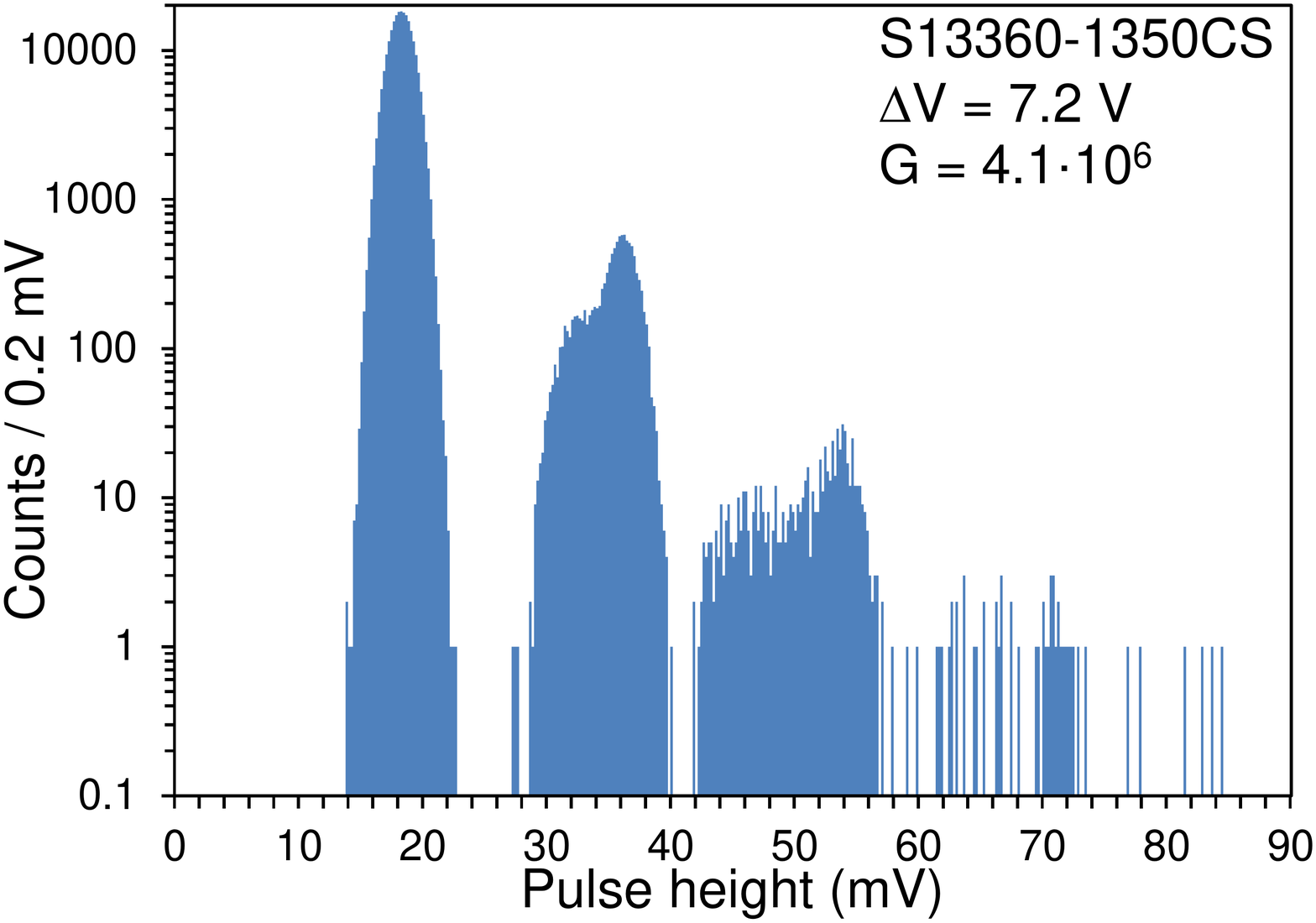}\,\,
\includegraphics[width=.45\textwidth]{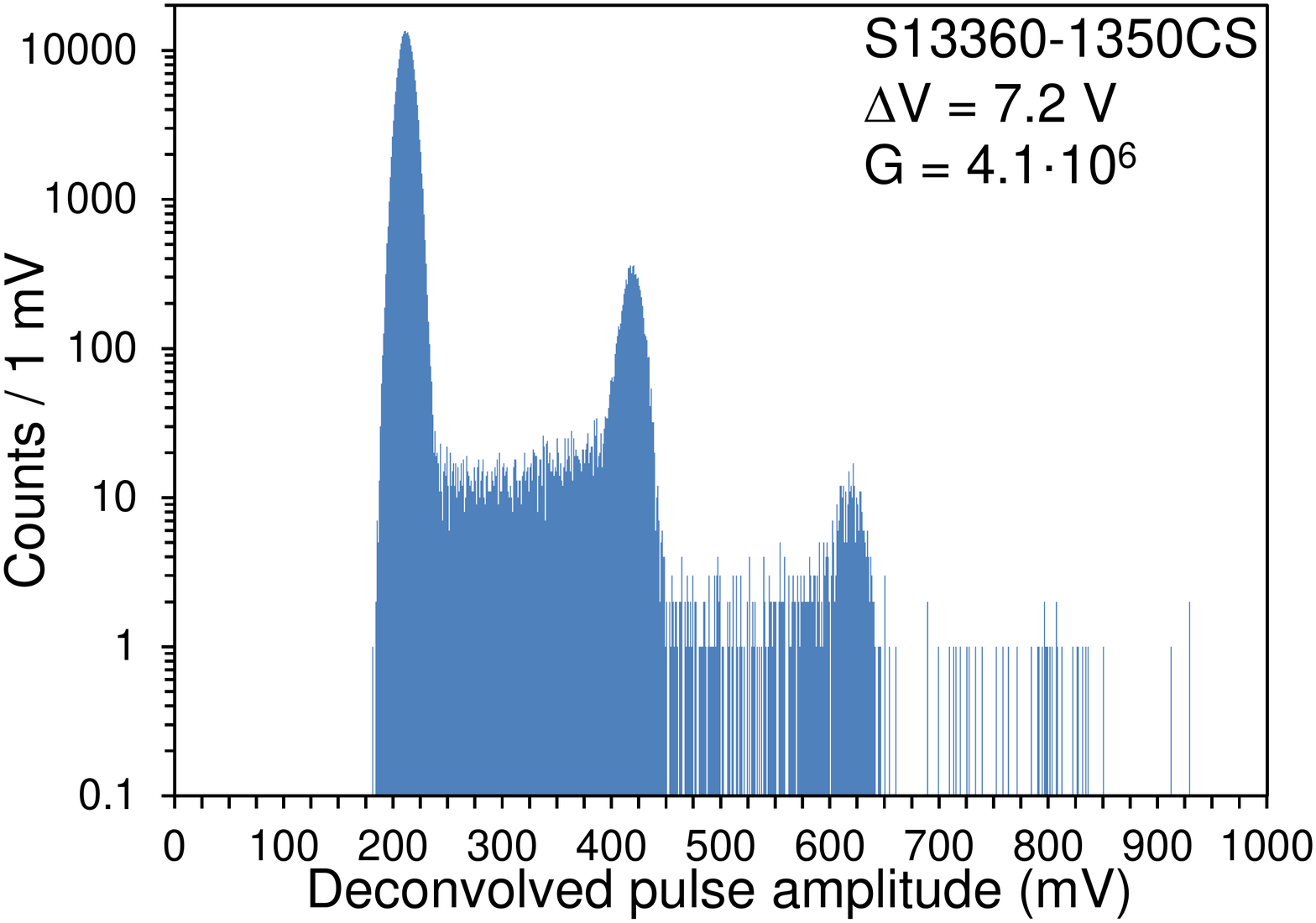}
\caption{Pulse height (left) and deconvolved pulse amplitude (right) spectra obtained for the S13360-1350CS MPPC at dark conditions.
The probability of prompt crosstalk is much smaller than in the S12571-050C MPPC despite the higher gain.
The peaks are asymmetrical because some crosstalk avalanches are not perfectly synchronous with the primary one.
This effect is much more noticeable in the deconvolved signal because of the shortening of pulses.}
\label{fig:spectra_13360}
\end{figure}

The spectra obtained for the S10362-11 and S12571 series show clear Gaussian peaks
(figure~\ref{fig:spectrum_height_12571}). Thus, a multi-Gaussian fit was performed to correct for peak overlapping for
those MPPCs. On the other hand, for the new S13360 series, with trenches, the spectra exhibit asymmetrical peaks for
crosstalk multiplicity $\geq1$, especially when using the deconvolved pulse amplitude (figure~\ref{fig:spectra_13360}).
This is due to pulses formed by two or more avalanches adding up in a time interval less than 3~ns, since these events
have to pass the cut of piled-up pulses, but asynchronously enough to give a pulse amplitude lower than the expected
one for perfectly synchronous avalanches. The above-mentioned spike component of the pulses enhances this effect, which
is much more noticeable in the deconvolved signal because of the shortening of pulses. The trenches clearly reduce the
prompt crosstalk, but this lack of synchronism suggests that a significant fraction of the prompt crosstalk remaining
is induced by minority carriers generated at very low depth in the substrate, that is, close to the depletion region of
neighboring pixels, as will be demonstrated in section~\ref{sec:MC} by Monte Carlo simulation. Coincidences of
uncorrelated pulses cannot justify it, because the dark count rate is extremely low.

\subsection{Overall prompt crosstalk probability}\label{ssec:CT_results}

The overall prompt crosstalk probability $\varepsilon$ versus gain is shown in figure~\ref{fig:promptCT} for most of
the detectors that were characterized. The uncertainties, which are basically statistical (see~\cite{Gallego} for
details), are less or around 1\% at moderate and high gain, except for the S10362-11-025C, S13360-1325CS and
S13360-3025CS MPPCs, for which the uncertainties are around 10\%. It can be seen that $\varepsilon$ grows more than
linearly with gain. The number of emitted secondary photons is proportional to the avalanche carrier multiplication,
but the probability of a photon to trigger an avalanche in a neighboring pixel also increases with overvoltage. On the
other hand, $\varepsilon$ must saturate at high gain. According to that, the following function was fitted to data:

\begin{figure}[t]
\centering
\includegraphics[width=.6\textwidth]{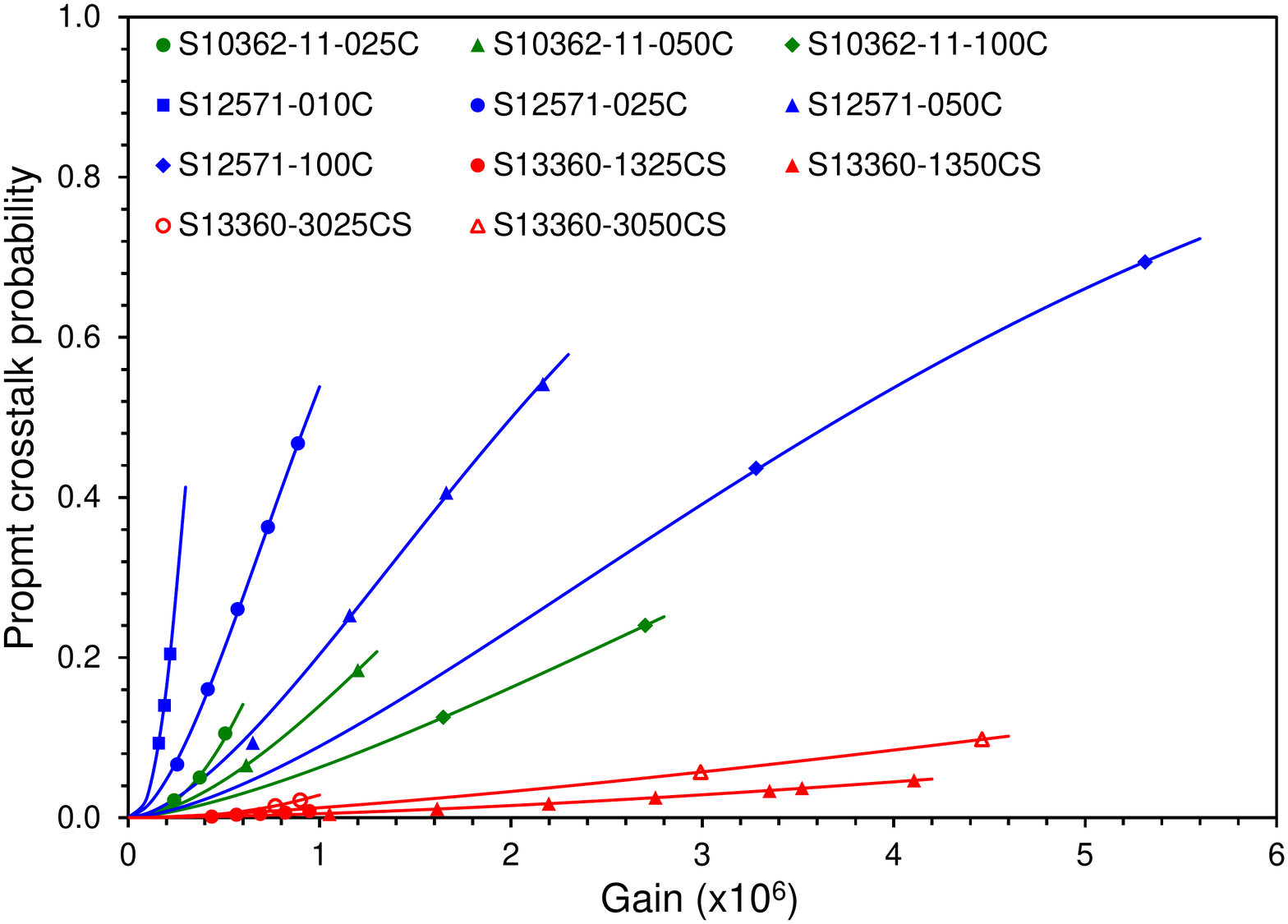}
\caption{Overall prompt crosstalk probability as a function of gain for various MPPCs.
Symbols represent experimental data and lines represent the fits of eq.~(\protect\ref{fit_CT}) to data.}
\label{fig:promptCT}
\end{figure}

\begin{equation}
\varepsilon(X)=1-\exp\left[-\left(\frac{X}{K}\right)^{(1+\alpha)}\right]\,,
\label{fit_CT}
\end{equation}
where $X$ can represent either the gain or the overvoltage, and $K$ is a MPPC constant in the corresponding units. The
index $\alpha$ is positive and accounts for the increase of the trigger probability with overvoltage assuming a power
law. In principle, the overvoltage dependence of the trigger probability can be extracted from the curve of the photon
detection efficiency (PDE) versus overvoltage, which is provided by Hamamatsu. This curve can be described by a power
law with exponent around 0.5, but with a positive overvoltage offset that increases with decreasing pixel size.
Nonetheless, the above function fits well to data, as shown in the figure.

The fitted $K$ and $\alpha$ values, as well as the proportionality constant between gain $G$ and overvoltage $\Delta
V$, are listed in table~\ref{tab:fit_CT}. The $K$ and $\alpha$ parameters are correlated with each other and therefore
their uncertainties may be large. Nevertheless, the fitted $\varepsilon$ values agree with the measured ones within
experimental uncertainties.

\begin{table}[!t]
\caption{Proportionality constant between gain $G$ and overvoltage $\Delta V$, and the fitted $K$ and $\alpha$ values
to prompt crosstalk data for all the MPPCs that were characterized.}%
\label{tab:fit_CT}%
\smallskip%
\centering%
\begin{tabular}{|cccc|}
\hline
MPPC & $\frac{{\rm d}G}{{\rm d}\Delta V}$ ($\times 10^6/V$) & $K$ (V) & $\alpha$\\
\hline
S10362-11-025C & 0.133 &  10.57 & 1.20 \\
S10362-11-050C & 0.580 &   5.40 & 0.65 \\
S10362-11-100C & 2.360 &   2.79 & 0.45 \\
S10362-33-100C & 2.360 &   2.56 & 0.62 \\
\hline
S12571-010C    & 0.030 &  12.81 & 1.71 \\
S12571-025C    & 0.158 &   7.37 & 0.71 \\
S12571-050C    & 0.506 &   4.98 & 0.60 \\
S12572-050C    & 0.506 &   4.03 & 0.56 \\
S12571-100C    & 2.068 &   2.30 & 0.52 \\
\hline
S13360-1325CS  & 0.129 & 103.24 & 0.82 \\
S13360-3025CS  & 0.129 &  32.23 & 1.50 \\
S13360-1350CS  & 0.571 &  49.38 & 0.58 \\
S13360-3050CS  & 0.571 &  39.42 & 0.40 \\
\hline
\end{tabular}
\end{table}

Our results are summarized in the following points:

\begin{enumerate}
\item The prompt crosstalk increases rapidly as the pixel size decreases keeping the same gain. This is because
    secondary photons have more chances to escape a small pixel.

\item The $\alpha$ index also increases with decreasing pixel size. The cause is not clear, but this variation is
    consistent with the above-mentioned shift of the PDE curve towards positive overvoltages.

\item The S12571 series have a prompt crosstalk probability higher than the old S10362-11 series, especially for
    the 25~$\mu$m pixel pitch type. In fact, Hamamatsu reports a fill factor of 65\% for the S12571-025C MPPC in
    contrast to 30.8\% for the S10362-11-025C MPPC, although the fill factors of the 50$\mu$m and 100~$\mu$m pixel
    pitch types are very similar in both series.

\item The prompt crosstalk probability for the S13360-1325CS and S13360-1350CS MPPCs is lower by a factor of around
    80 and 35, respectively, than for the S12571-025C and S12571-050C ones.

\item The 3~mm~$\times$~3~mm active area MPPCs have somewhat higher prompt crosstalk probabilities than their small
    active area counterparts, which have a lower number of pixels.
\end{enumerate}

\subsection{Prompt crosstalk multiplicity}\label{ssec:CT_multiplicity}

\begin{figure}[t]
\centering
\includegraphics[width=.45\textwidth]{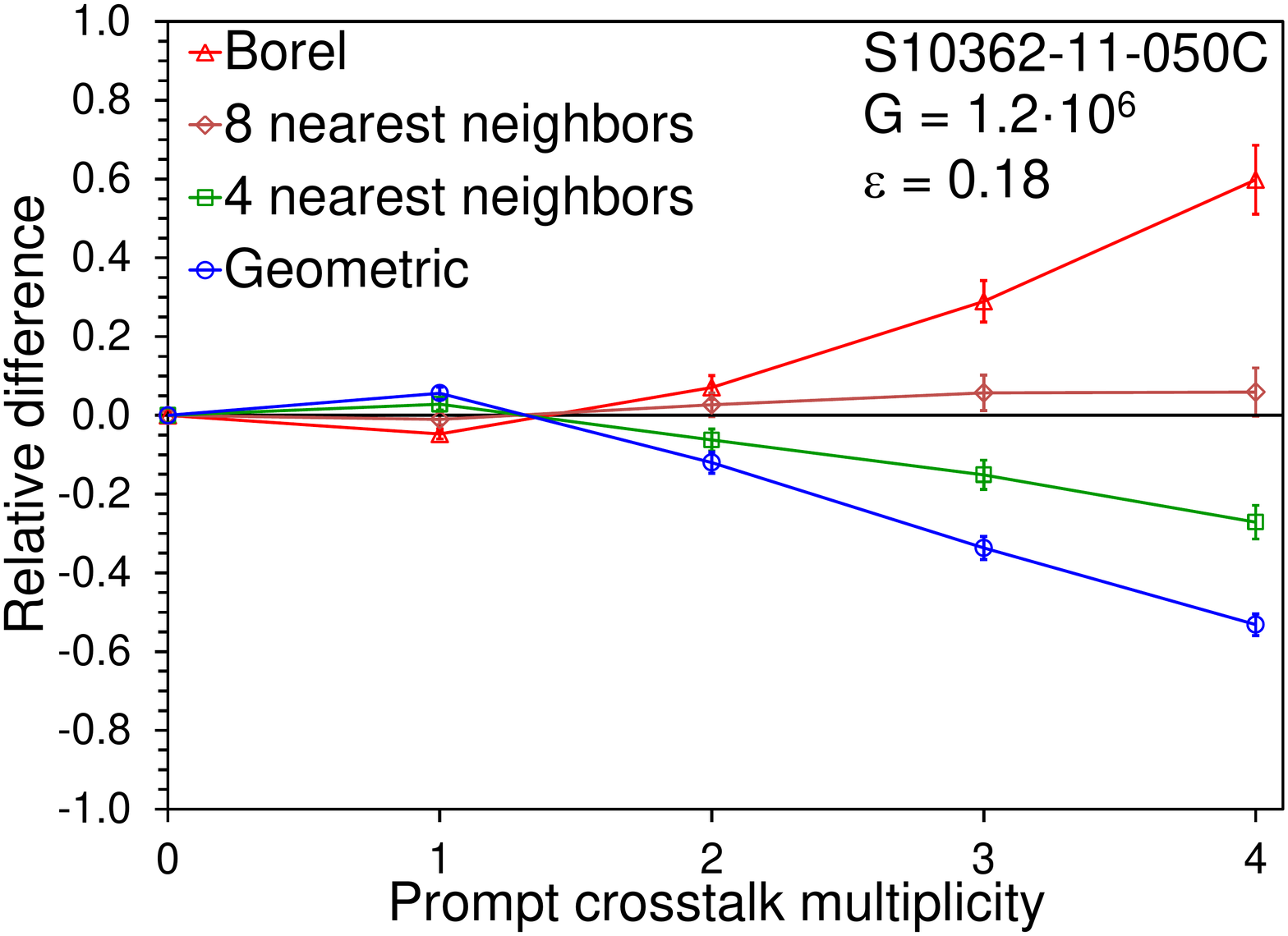}\,\,
\includegraphics[width=.45\textwidth]{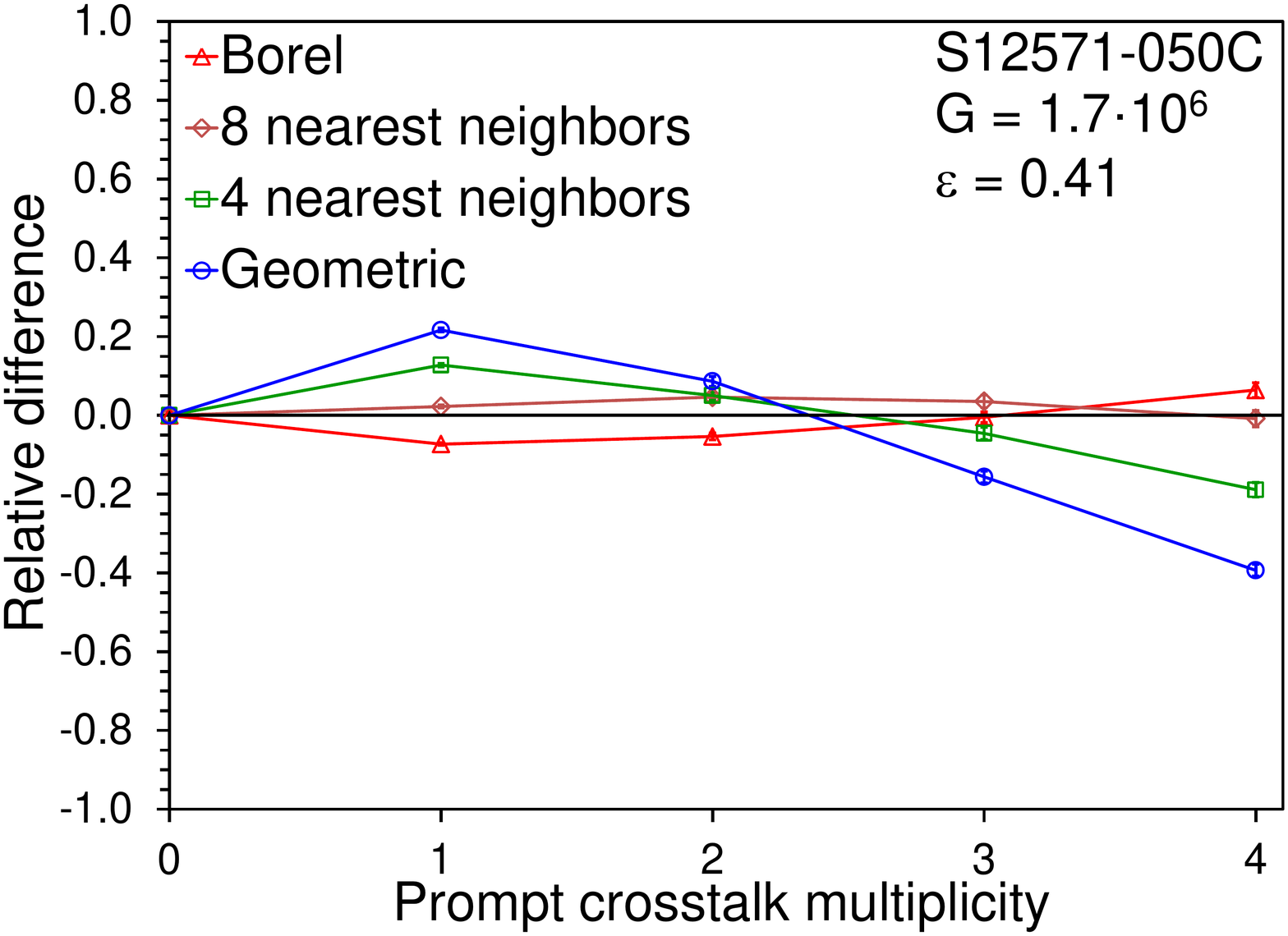}
\includegraphics[width=.45\textwidth]{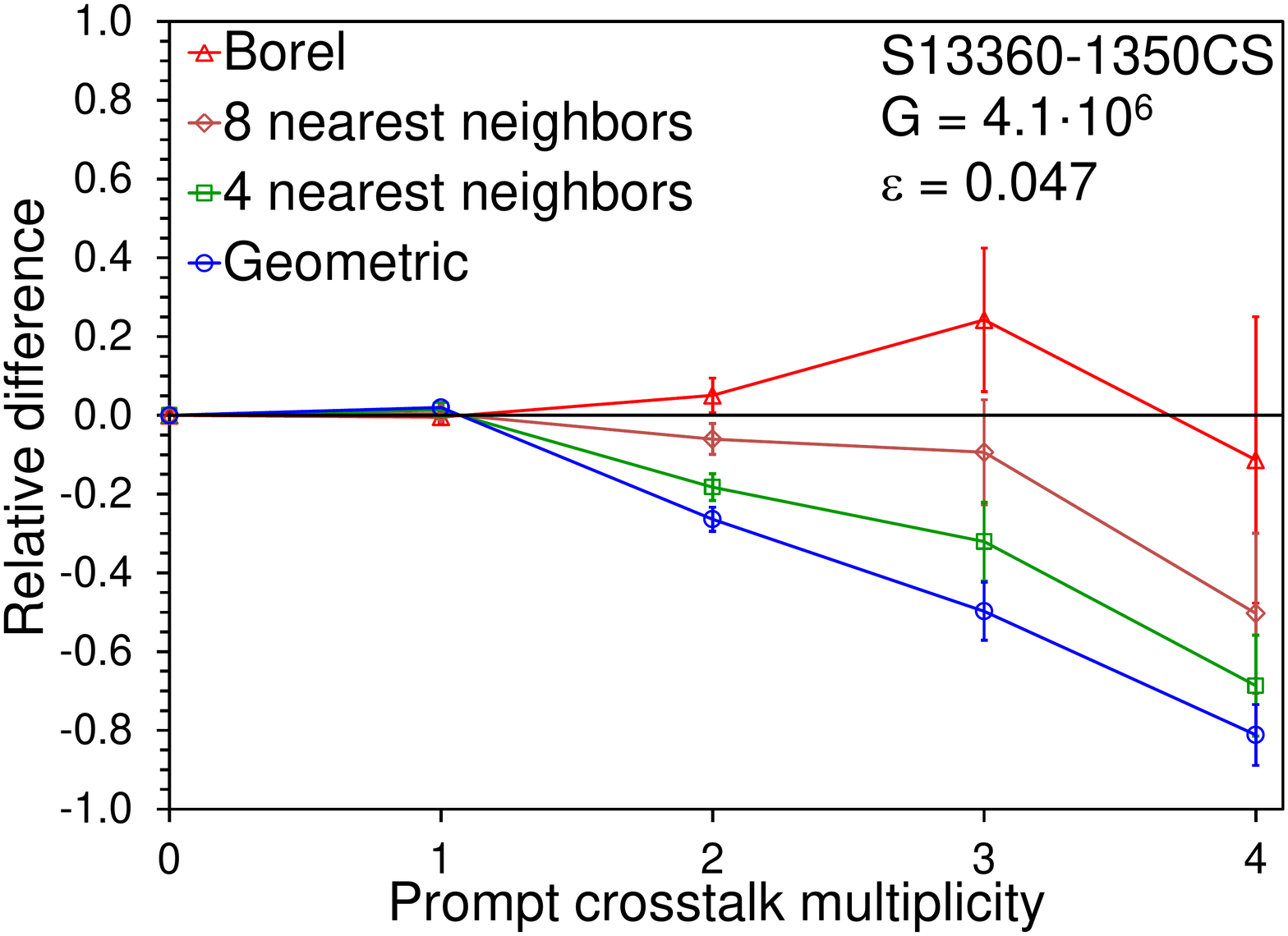}
\caption{Relative differences of the prompt crosstalk multiplicity distributions predicted by different analytical models
with respect to the experimental one for 50~$\mu$m pixel pitch MPPCs.
Results suggest that prompt crosstalk affects at least the 8 nearest neighboring pixels, but with a higher probability for the 4 nearest ones.
Uncertainties are large for the S13360-1350CS MPPC because the prompt crosstalk probability is very low.}
\label{fig:comparison_CT}
\end{figure}

The probability distribution of the prompt crosstalk multiplicity for a given $\varepsilon$ value depends on the number
of pixels that can be reached by secondary photons from the primary pixel and on crosstalk cascading, i.e., secondary
avalanches inducing further avalanches. This was modeled in a previous work~\cite{Gallego}, where exact analytical
formulae were derived from simple geometrical assumptions. For the MPPCs that were characterized in the present work,
two models were found to fit data in general, namely that either the 4 nearest neighbors or the 8 nearest neighbors of
each fired pixel have equal probability of prompt crosstalk, whereas the probability is zero in the remaining pixels.
In figure~\ref{fig:comparison_CT}, we show the relative difference of the multiplicity distributions predicted by both
models with respect to the experimental one for 50~$\mu$m pixel pitch MPPCs and up to multiplicity 4. The model
predictions were normalized so that the probability of multiplicity 0 equals $1-\varepsilon$. Also shown in the figures
are the results for two other analytical models proposed in~\cite{Vinogradov}, which represent limit situations
regarding the number of neighboring pixels subject to prompt crosstalk (see~\cite{Gallego} for details). An empirical
formula proposed in~\cite{FACT} proved to describe the crosstalk multiplicity distribution for the S10362-33-050C MPPC,
but we did not included it in the comparison because it uses a fitting shape parameter and has no clear geometrical
interpretation.

The following features were observed in the comparison of data with the model predictions:

\begin{enumerate}
\item For a given MPPC series, the model that agrees with data the best is independent of the bias voltage and the
    size of the active area (see also~\cite{Rosado}). This indicates that the prompt crosstalk is mostly determined
    by the pixel structure.

\item The larger the pixel size, the lower the number of neighboring pixels subject to prompt crosstalk according
    to our models. Although this trend is subtle, it is the expected one for an exponential attenuation of
    secondary photons. Small differences were also observed between the S10362-11 and S12571 series for the same
    pixel pitch.

\item The agreement of models with data for the S13360 series is compatible with the other series. However, the
    uncertainties are large due to the low probability of prompt crosstalk and no conclusions can be drawn.
\end{enumerate}

\section{Characterization of afterpulsing and delayed crosstalk}\label{sec:aft}

\subsection{Procedure}\label{ssec:procedure}

In a previous work~\cite{Rosado}, we presented a procedure to characterize the afterpulsing and delayed crosstalk
through their delay time and amplitude distributions. The procedure has been upgraded and it is described in detail
here.

\begin{figure}[t]
\centering
\includegraphics[width=.8\textwidth]{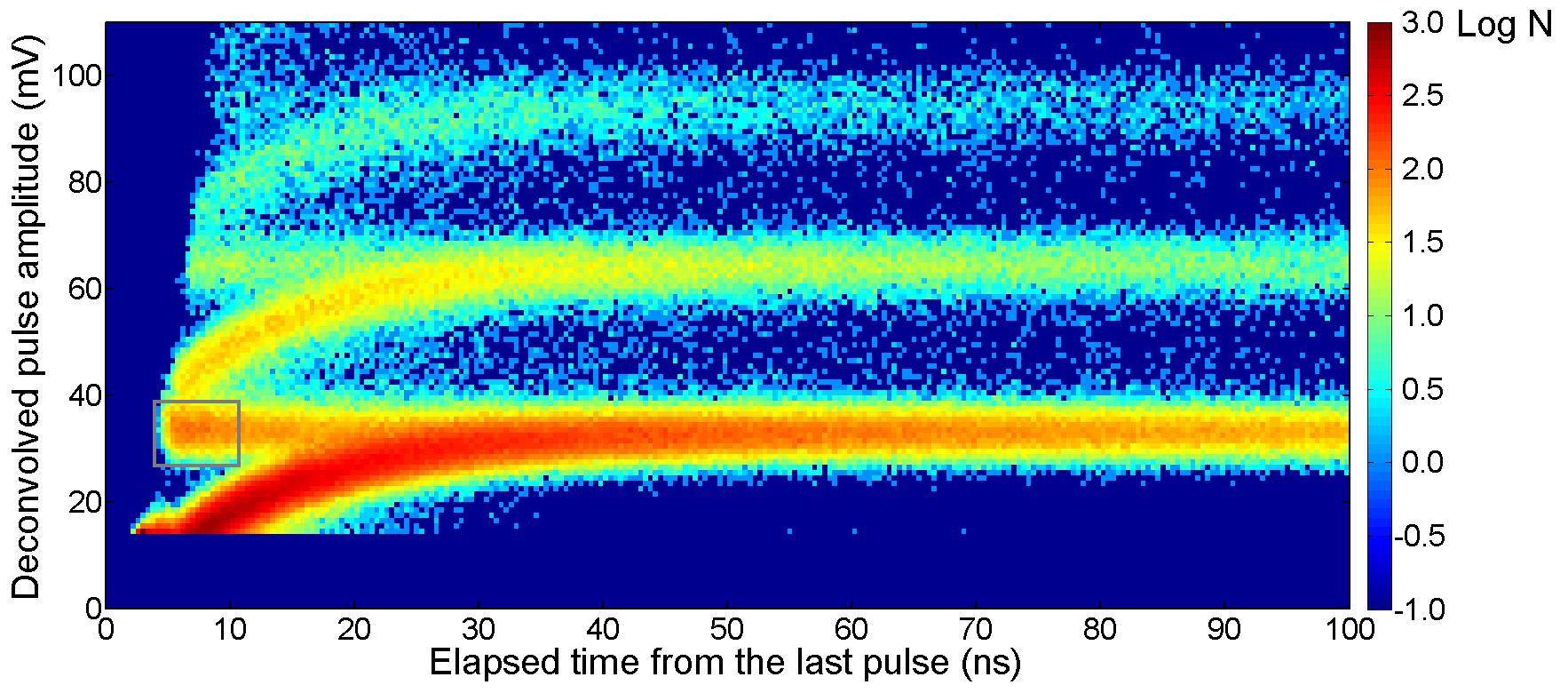}
\caption{2D histogram of deconvolved pulse amplitude versus elapsed time from the last pulse obtained for the S10362-11-050C MPPC
at an overvoltage of 2.1~V ($G=1.2\cdot10^6$).
Data are distributed in several horizontal bands corresponding, from bottom to top, to prompt crosstalk multiplicity 0, 1, 2,~...
The low-amplitude branch of each band is made up by afterpulses.
The grey square indicates the secondary pulses selected to obtain the distribution $P^0_{\rm no AP}(t)$.}
\label{fig:amp_time_10362}
\end{figure}

The first step was to obtain the 2D histogram of the deconvolved pulse amplitude versus the elapsed time from the last
pulse for each MPPC and bias voltage. For the sake of clarity, a pulse contributing to this histogram will be referred
to hereafter as a secondary pulse, and its preceding pulse giving the origin of time, as the primary one, although they
are not necessarily correlated with each other. To obtain the histogram, we only selected secondary pulses of which the
primary pulse has an amplitude within the range of values corresponding to a single avalanche (i.e., no prompt
crosstalk) and passes the cut of piled-up pulses. The primary pulse was also imposed to be separated by at least 500~ns
from previous ones to prevent contamination of secondary pulses correlated with them. An example of such a histogram
for the S10362-11-050C MPPC is shown as a color map in figure~\ref{fig:amp_time_10362}. The undermost horizontal band
is made up by secondary pulses without prompt crosstalk, and the upper bands correspond to prompt crosstalk
multiplicities equal or greater than 1. The low-amplitude branch of each band is due to afterpulses that are produced
while the pixel is still recovering from the primary pulse. Occurrence of two consecutive uncorrelated dark events in
the same pixel also contributes to this branch, but it is very unlikely. Full-amplitude pulses at short time are
delayed-crosstalk and uncorrelated dark events (in different pixels).

\begin{figure}[t]
\centering
\includegraphics[width=.6\textwidth]{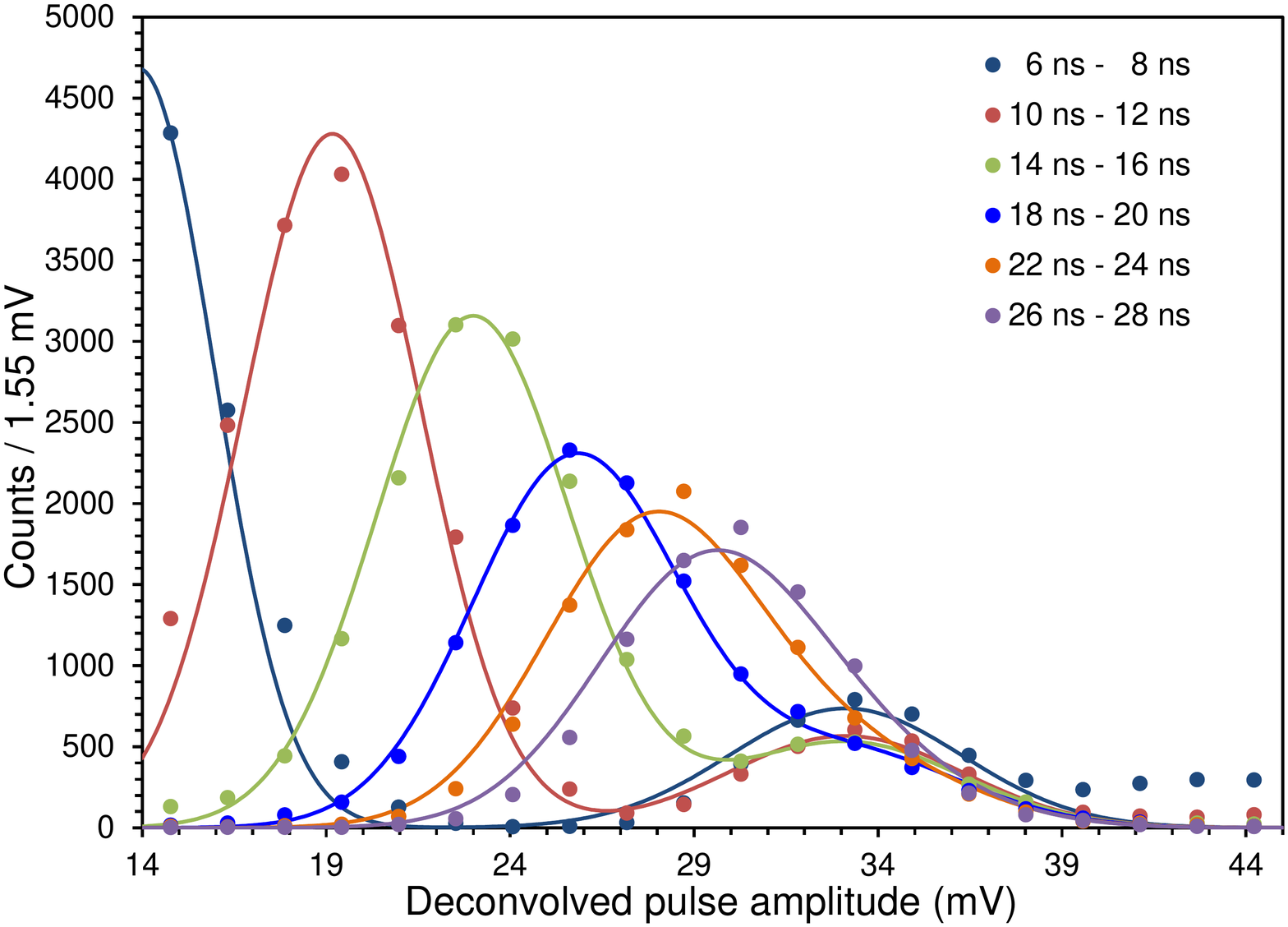}
\caption{Pulse amplitude distributions extracted from the 2D histogram of figure~\protect\ref{fig:amp_time_10362} by integrating in 2~ns intervals.
The distributions are restricted to single-avalanche pulses (no prompt crosstalk).
A double-Gaussian fit was performed to characterize the pixel recovery and the losses of afterpulses with amplitude below the discrimination threshold
($A_{\rm thr}=14$~mV in this case).}
\label{fig:recovery_10362}
\end{figure}

In a second step, the amplitude-time correlation of afterpulses was utilized to characterize the recovery function of
the MPPC. The 2D histogram was integrated in intervals of 2~ns limited to the first band of secondary pulses, obtaining
a collection of amplitude histograms, each one at a different stage of the pixel recovery. This is shown in
figure~\ref{fig:recovery_10362}, where two Gaussian-like components, corresponding to afterpulses and full-amplitude
pulses, are distinguished for every time interval. We observed that the variance of both Gaussian components increases
linearly with increasing gain, as expected from the Poissonian nature of the avalanche carrier multiplication. Thus, a
double-Gaussian fit was performed with the following restrictions: a)~the mean of the component of full-amplitude
pulses was set equal to the mean value $A_1$ of the first peak in the deconvolved pulse amplitude spectrum described in
section~\ref{ssec:spectra}, b)~the variance of both Gaussian components was forced to be proportional to the mean,
where the proportionality constant $C$ was taken as a common fitting parameter for all the time intervals, and c)~the
mean of the component of afterpulses was assumed to vary with the time interval as

\begin{equation}
A(t_k)=A_1\cdot\theta(t_k-t_0)\cdot\left[1-\exp\left(-\frac{t_k-t_0}{t_{\rm rec}}\right)\right]\,,
\label{mean_amplitude}
\end{equation}
where $t_k$ is the central value of the $k$-th interval, $\theta(t)$ is the Heaviside step function, $t_{\rm rec}$ is
the recovery time and $t_0$ is a positive time offset that was needed to fit data for some MPPCs. This offset could
arise from the fact that the pixel voltage may drop below the breakdown voltage~\cite{Otono}. As expected, the fitted
$t_{\rm rec}$, $C$ and $t_0$ values showed no dependence on bias voltage. Average $t_{\rm rec}$ and $t_0$ values for
every MPPC are shown in table~\ref{tab:AP_results}.

\begin{figure}[t]
\centering
\includegraphics[width=.45\textwidth]{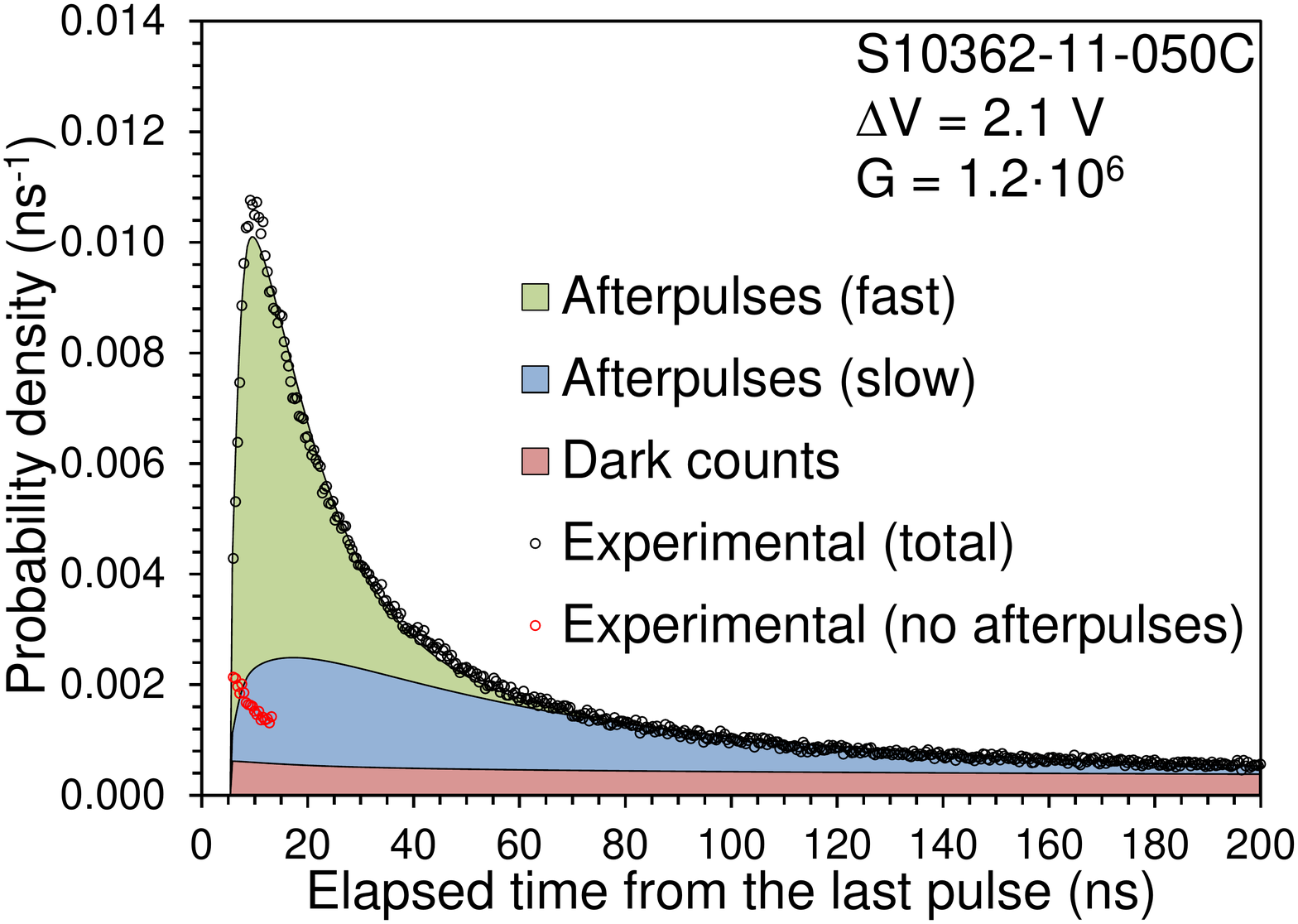}
\includegraphics[width=.45\textwidth]{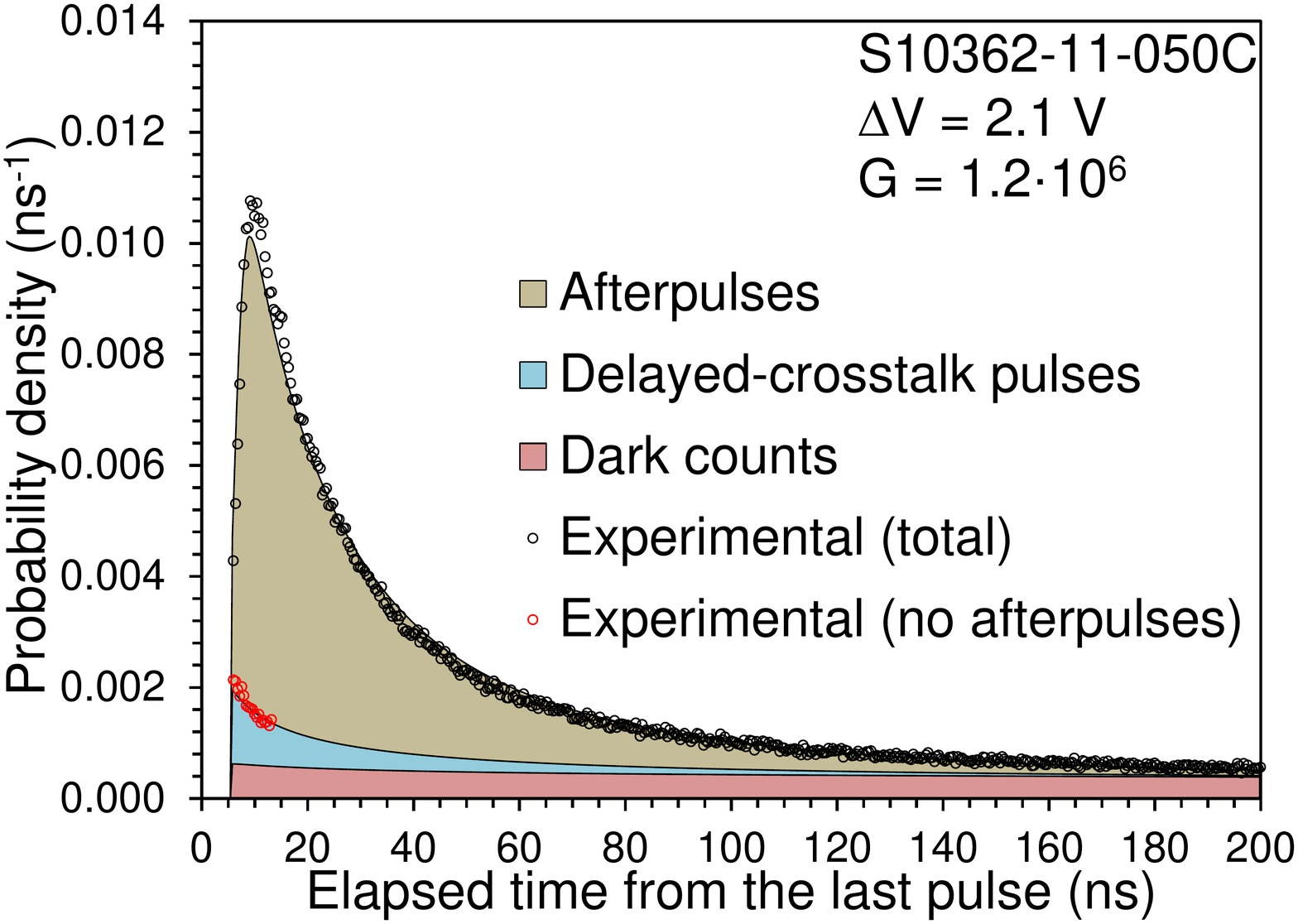}
\caption{Delay time distribution of secondary pulses for the S10362-11-050C MPPC extracted from figure \protect\ref{fig:amp_time_10362}.
In the left-hand plot, two components of trap-assisted afterpulses with different lifetimes were fitted to data.
The fit is good, but the high density of full-amplitude pulses at short time (red open circles) cannot be justified by uncorrelated dark counts only.
In the right-hand plot, only the components of afterpulsing and delayed crosstalk due to minority carriers diffusing in the substrate were fitted to data.
Both the $P(t)$ and $P_{\rm no AP}(t)$ distributions are properly described by this model.}
\label{fig:time_dist_10362}
\end{figure}

\begin{figure}[t]
\centering
\includegraphics[width=.45\textwidth]{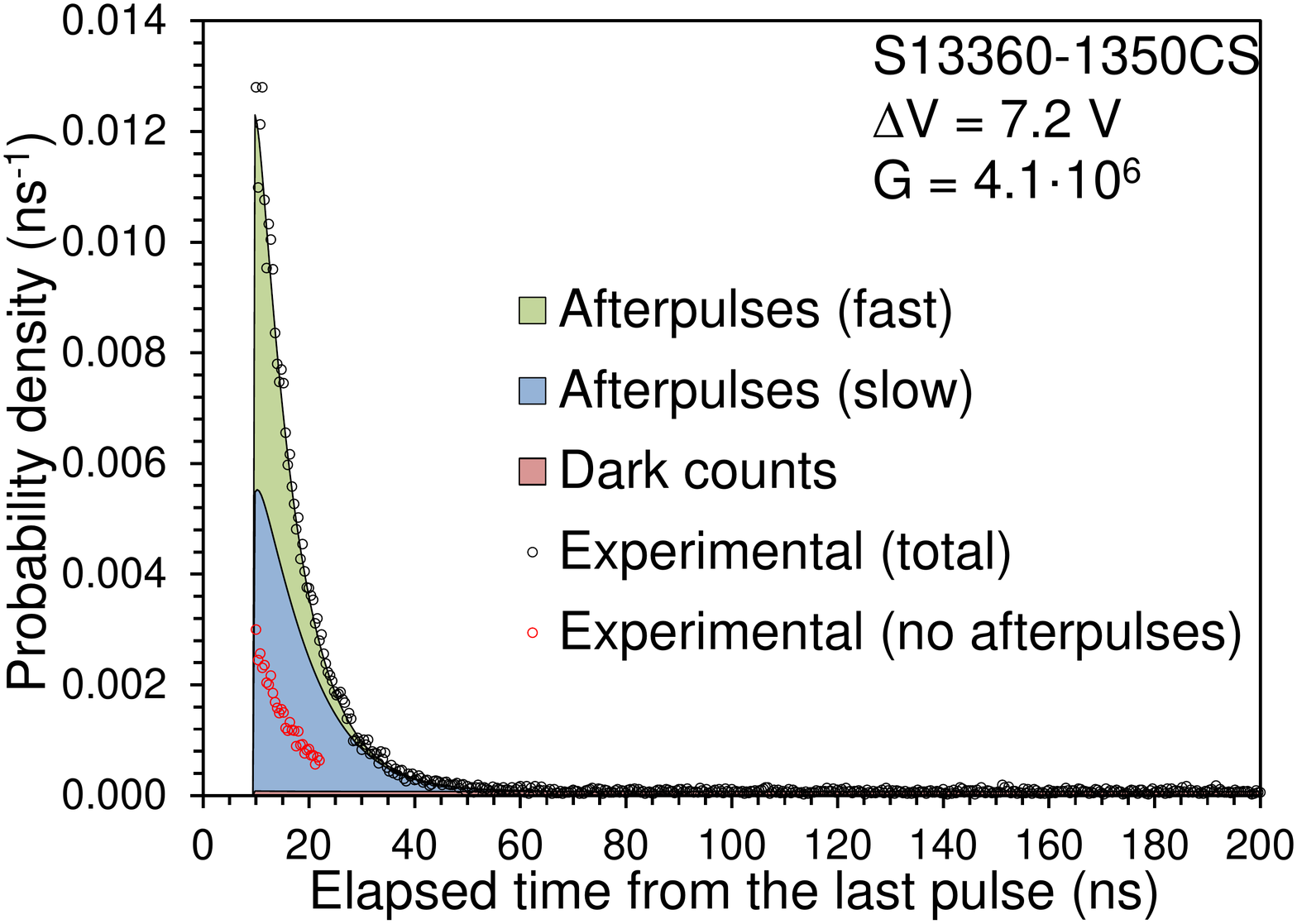}
\includegraphics[width=.45\textwidth]{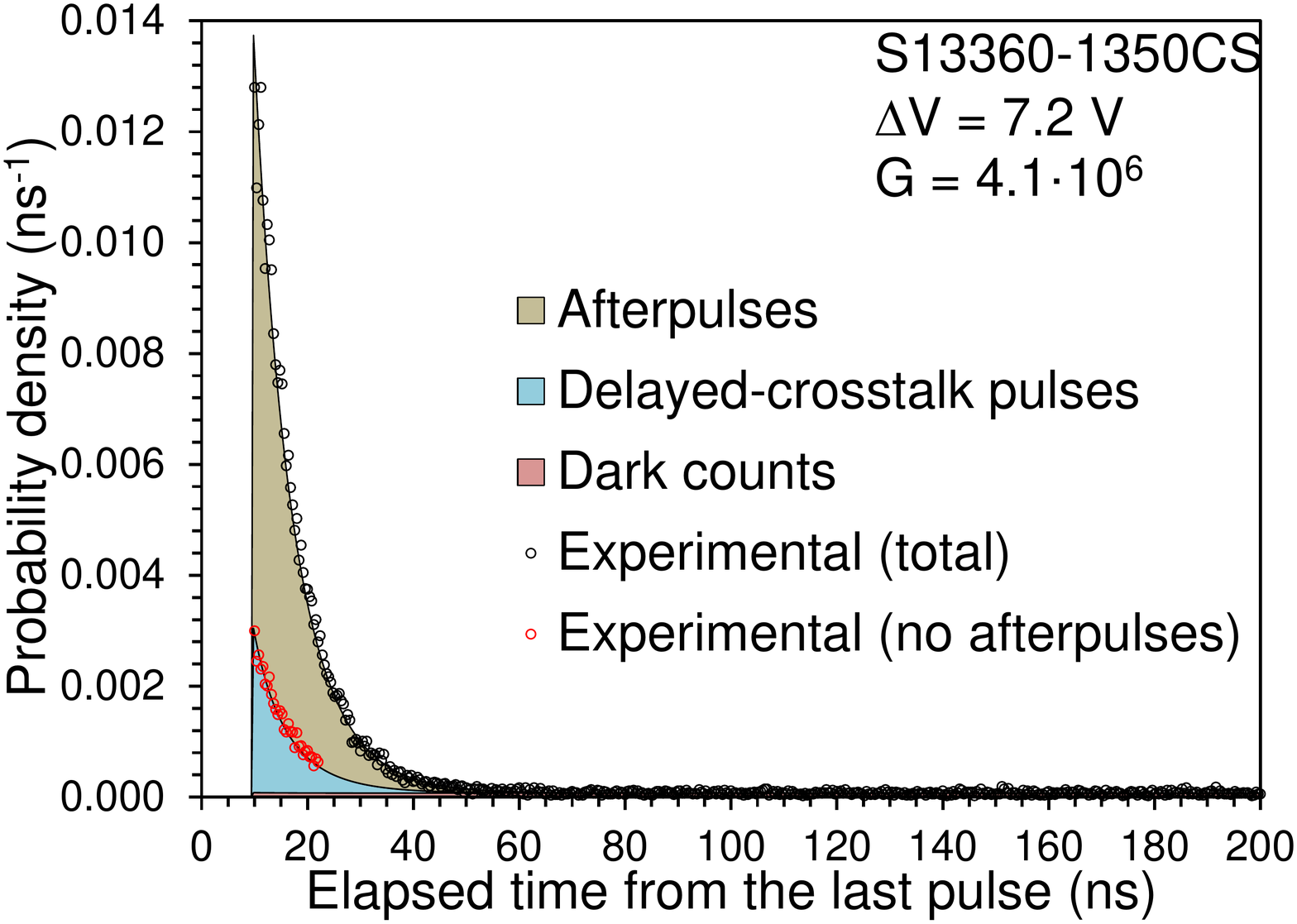}
\caption{The same as figure \protect\ref{fig:time_dist_10362} for the S13360-1350CS MPPC.
The correlated noise decays much faster than for the S10362-11-050C MPPC.}
\label{fig:time_dist_13360}
\end{figure}

In a third step, the 2D histogram was integrated over amplitude and normalized to obtain the probability density
distribution of delay time $P(t)$, including all types of the secondary pulses. Although some pulses with very short
delay time were able to be resolved, we set a minimum delay time $t_{\rm min}$ that defined the effective time
resolution ($t_{\rm min}$ was either 6~ns or 10~ns, depending on the MPPC). The resulting $P(t)$ distribution is
represented by black open circles in figures~\ref{fig:time_dist_10362} and~\ref{fig:time_dist_13360} for the
S10362-11-050C and S13360-1350CS MPPCs, respectively. In addition, we obtained the probability density distribution
$P_{\rm no AP}(t)$ where the afterpulses were removed for delay time shorter than a certain value $t_{\rm max}$
depending on the MPPC (red open circles in figures~\ref{fig:time_dist_10362} and~\ref{fig:time_dist_13360}). To do
that, we first obtained a distribution $P^0_{\rm no AP}(t)$ selecting only full-amplitude pulses of the undermost band
of the 2D histogram, that is, delayed-crosstalk and uncorrelated dark events without prompt crosstalk, as illustrated
by the grey square in figure~\ref{fig:amp_time_10362}. Then, the $P_{\rm no AP}(t)$ distribution was estimated by

\begin{equation}
P_{\rm no AP}(t)=(1+\varepsilon)\cdot P^0_{\rm no AP}(t)\,,
\label{corr_CT_secondaries}
\end{equation}
where the factor $1+\varepsilon$ adds the probability of secondary pulses with prompt crosstalk, but still excluding
afterpulses with or without prompt crosstalk.

The last step was to find an appropriate combination of noise components that describes the experimental distributions
$P(t)$ and $P_{\rm no AP}(t)$. The $P(t)$ distribution can be expressed as

\begin{equation}
P(t)=\exp\left[-R_{\rm DC}\cdot(t-t_{\rm min})-\sum_i{\int_{t_{\rm min}}^t\!\!\!\!\!f_i(s)\cdot{\rm d}s}\right]\cdot
\left[R_{\rm DC}+\sum_i{f_i(t)}\right]\,,
\label{prob_secondaries}
\end{equation}
where $R_{\rm DC}$ is the dark count rate and $f_i(t)$ is the density distribution of secondary pulses of type $i$
directly correlated with the primary pulse, i.e., excluding cascades. To derive the above equation, we assumed that the
number of secondary pulses in a given time interval follows a Poisson distribution. Likewise, the $P_{\rm no AP}(t)$
distribution is expressed as

\begin{equation}
P_{\rm no AP}(t)=\exp\left[-R_{\rm DC}\cdot(t-t_{\rm min})-\int_{t_{\rm min}}^t\!\!\!\!\!f_{\rm CT}(s)\cdot{\rm d}s\right]\cdot
\left[R_{\rm DC}+f_{\rm CT}(t)\right]\,,
\label{prob_no_AP}
\end{equation}
where $f_{\rm CT}(t)$ is the density distribution of delayed-crosstalk pulses.

Note that $P(t)$ is proportional to $\exp(-R_{\rm DC}\cdot t)$ at long delay time, because the correlated noise
extinguishes with time. We used it to determine $R_{\rm DC}$ by means of an exponential fit for delay time greater than
1~$\mu$s. As expected, the fitted $R_{\rm DC}$ values were proportional to overvoltage and were lower than 1~MHz in
most of the cases, with the new MPPC series having the lowest $R_{\rm DC}$ values. Even if a $\mu$s-life component of
correlated noise were present, it would be undistinguishable from the uncorrelated dark noise.

To model the correlated noise components, we proceeded in the following way. Each component of trap-assisted
afterpulses was parameterized by

\begin{equation}
f_{\rm trap}(t)=C_{\rm trap}\cdot\exp\left(-\frac{t}{\tau_{\rm trap}}\right)\cdot\mu(t)\cdot k(t)\,,
\label{f_AP_traps}
\end{equation}
where the coefficient $C_{\rm trap}$ only depends on overvoltage and $\tau_{\rm trap}$ is the trap lifetime. The factor
$\mu(t)$ accounts for the recovery of the trigger probability in the primary pixel and it is assumed to be

\begin{equation}
\mu(t)=\frac{A(t)}{A_1}\,,
\label{mu}
\end{equation}
where $A(t)$ is given by (\ref{mean_amplitude}). The factor $k(t)$ is the fraction of afterpulses that surpass the
discrimination threshold $A_{\rm thr}$ at a given delay time. Note that afterpulses with prompt crosstalk have always
amplitudes above $A_{\rm thr}$. Taking into account that the probability of an avalanche to induce prompt crosstalk is
proportional to its carrier multiplication, the time-varying fractions of afterpulses with and without prompt crosstalk
can be estimated by $\varepsilon\cdot\mu(t)$ and $1-\varepsilon\cdot\mu(t)$, respectively. Now, let $k_0(t)$ be the
fraction of afterpulses without prompt crosstalk that surpass the threshold. Assuming a Gaussian distribution of
afterpulse amplitude at given delay time, as explained above, $k_0(t)$ is

\begin{equation}
k_0(t)=\int_{A_{\rm thr}}^{\infty}\frac{1}{\sqrt{2\cdot\pi\cdot C\cdot A(t)}}\cdot\exp\left[-\frac{\left(x-A(t)\right)^2}{2\cdot C\cdot A(t)}\right]\cdot{\rm d}x\,,
\label{k_0}
\end{equation}
where $C$ is already known. Therefore, $k(t)$ is given by

\begin{equation}
k(t)=\left(1-\varepsilon\cdot\mu(t)\right)\cdot k_0(t)+\varepsilon\cdot\mu(t)\,.
\label{k}
\end{equation}

We also considered the components of afterpulsing and delayed crosstalk induced by minority carriers diffusing in the
substrate. As will be explained in section~\ref{sec:MC}, a Monte Carlo simulation was performed to evaluate the
diffusion time distribution for minority carriers reaching either the primary pixel or any other pixel. In both cases,
the distribution was found to be well described by the product of a power law and an exponential decay function related
to the minority carrier recombination. Therefore, we obtained

\begin{equation}
f_{\rm AP}(t)=C_{\rm AP}\cdot t^a\cdot\exp\left(-\frac{t}{\tau_{\rm bulk}}\right)\cdot\mu(t)\cdot k(t)\,,
\label{f_AP_diff}
\end{equation}
and

\begin{equation}
f_{\rm CT}(t)=C_{\rm CT}\cdot t^b\cdot\exp\left(-\frac{t}{\tau_{\rm bulk}}\right)\,,
\label{f_CT_diff}
\end{equation}
where the $a$ and $b$ exponents are both negative and $\tau_{\rm bulk}$ is the minority carrier lifetime in the silicon
substrate, also called bulk lifetime. Notice that the delayed crosstalk is not affected by the $\mu(t)$ and $k(t)$
factors.

The number of trap-assisted afterpulse components, if any, is not known a priori. Besides, many different combinations
of noise components may fit to the experimental time distributions. Therefore, we opted for making two simple
hypotheses. In the first one (H1), we considered exclusively two trap-assisted afterpulse components: ``slow'' and
``fast'' (see, e.g.,~\cite{Vacheret}); four fitting parameters were defined: $C_{\rm slow}$, $\tau_{\rm slow}$, $C_{\rm
fast}$ and $\tau_{\rm fast}$; and the $P_{\rm no AP}(t)$ distribution was not used in the fit. In the second one (H2),
only afterpulsing and delayed crosstalk related to carrier diffusion were considered; the fitting parameters were
$C_{\rm AP}$, $C_{\rm CT}$ and $\tau_{\rm bulk}$; we set $a=-1$ and $b=-0.5$, according to our simulation results (see
section~\ref{sec:MC}), to reduce the number of degrees of freedom; and we put the extra restriction that the sum of the
dark-count and delayed-crosstalk components has to fit to $P_{\rm no AP}(t)$. The results for both hypotheses are
discussed in the next section.

\subsection{Results}\label{ssec:AP_results}

The hypothesis H1 was found to describe successfully the shape of $P(t)$ for all the MPPCs and bias voltages. As an
examples, the best fits to $P(t)$ for the S10362-11-050C and S13360-1350CS MPPCs are shown in the left-hand plots of
figures~\ref{fig:time_dist_10362} and~\ref{fig:time_dist_13360}, respectively, where the stacked areas represent the
three components considered: dark counts and both types of trap-assisted afterpulses. However, H1 has two deficiencies
that make it incompatible with data. In the first place, a significant fraction of pulses that populate $P(t)$ at short
delay time are full-amplitude pulses that belong to $P_{\rm no AP}(t)$ and cannot be justified solely by dark counts.
It is clear that the delayed crosstalk is non-negligible. In the second place, the fitted $\tau_{\rm slow}$ and
$\tau_{\rm fast}$ values, which represent supposedly the lifetimes of two types of traps in the depleted region, vary
notably from MPPC to MPPC even within the same series. For instance, we obtained $\tau_{\rm slow}=70$~ns and $\tau_{\rm
fast}=10$~ns for the S10362-11-050C MPPC, but $\tau_{\rm slow}=7$~ns and $\tau_{\rm fast}=4$~ns for the S13360-1350CS
MPPC.

For the hypothesis H2, the fit to $P(t)$ was as good as for H1, despite the fact that the number of fitting parameters
was reduced from 4 to 3. In addition, a good fit to $P_{\rm no AP}(t)$ was achieved by including the component of
delayed crosstalk. The best fits for the S10362-11-050C and S13360-1350CS MPPCs are shown in the right-hand plots of
figures~\ref{fig:time_dist_10362} and~\ref{fig:time_dist_13360}, respectively. Note that the power-law factors in
equations~(\ref{f_AP_diff}) and (\ref{f_CT_diff}) properly account for the non-exponential tails of $P(t)$ and $P_{\rm
no AP}(t)$. The small inaccuracies of the fits ($<10\%$) are likely due to the assumptions made on the indexes $a$ and
$b$ as well as on the $\mu(t)$ and $k(t)$ functions, which determine the fast rising and the peak of $P(t)$.
Correlations were also found between all these parameters and the fitting parameters $C_{\rm AP}$, $C_{\rm CT}$ and
$\tau_{\rm bulk}$. Adding a trap-assisted afterpulse component would improve the fit, but this does not seem to be
necessary to describe the shape of $P(t)$. Some authors, who carried out a characterization of the afterpulsing and
delayed crosstalk in other SiPMs, attributed the afterpulses to trap-assisted processes only~\cite{Nagy,Acerbi}.
However, the occurrence of delayed crosstalk induced by minority carriers diffusing in the substrate should be along
with a component of afterpulsing. Consequently, we adopted the hypothesis H2 to interpret our data.

The fitted $\tau_{\rm bulk}$ values are a few hundreds of ns for the old S10362 series and around 10~ns for the newer
ones. This seems to be the main cause of the afterpulse reduction in the new MPPCs, since the lower the bulk lifetime
of the substrate, the lower the number of minority carriers reaching the depleted region of pixels. Using data
from~\cite{Wang}, these $\tau_{\rm bulk}$ values translate into an increase of the substrate doping density from
3~--~6~$\cdot10^{18}$~cm$^{-3}$ for the S10362 series to $2\cdot10^{19}$~cm$^{-3}$ for the newer ones. The recovery
time also increases from the old to the new series, contributing to the afterpulse reduction too. These results are
consistent with the findings of~\cite{Acerbi}.

\begin{table}[!t]
\caption{Fitting parameters of the model for afterpulsing and delayed crosstalk.
The integrals defined in eq.~(\protect\ref{N_sec}) calculated for $G=10^6$ are given in the last two columns.}%
\label{tab:AP_results}%
\smallskip
\footnotesize
\centering%
\begin{tabular}{|ccccccccc|}
\hline
MPPC           & $t_{\rm rec}$ (ns) & $t_0$ (ns) & $\tau_{\rm bulk}$ (ns) & $K_{\rm AP}$ (V) & $\beta $       & $\frac{C_{\rm CT}}{C_{\rm AP}}$ (ns$^{-0.5}$) & $N_{\rm AP}$ & $N_{\rm CT}$ \\
\hline
S10362-11-025C &  5.0               & 0.6        &  73.0                  & 11.78            & 1.40           & 0.060$^{\rm a}$                               & 0.74         &   0.25       \\
S10362-11-050C & 12.5               & 0.0        & 120.0                  &  4.86            & 1.05           & 0.021                                         & 0.25         &   0.04       \\
S10362-11-100C & 36.0               & 4.3        & 244.0                  &  2.26            & 0.67           & 0.015                                         & 0.09         &   0.02       \\
S10362-33-100C & 43.7               & 2.4        & 395.0                  &  1.88            & 1.47           & 0.017                                         & 0.04         &   0.01       \\
\hline
S12571-010C    &  4.4               & 0.0        &   6.4                  & 25.96            & 1.71           & 0.140$^{\rm b}$                               & 0.88         &   0.43       \\
S12571-025C    &  8.5               & 0.0        &  10.8                  & 11.33            & 1.15           & 0.060                                         & 0.15         &   0.05       \\
S12571-050C    & 20.5               & 1.0        &  12.3                  &  6.24            & 0.89           & 0.021                                         & 0.04         &   0.01       \\
S12572-050C    & 20.4               & 0.0        &   9.6                  &  5.36            & 1.00$^{\rm c}$ & 0.021$^{\rm d}$                               & 0.04         &   0.01       \\
S12571-100C    & 56.1               & 1.1        &  11.0                  &  2.04            & 0.55           & 0.006                                         & 0.01         & $<0.01$      \\
\hline
S13360-1325CS  & 16.7               & 0.0        &   9.4                  & 15.90            & 1.07           & 0.044                                         & 0.07         &   0.02       \\
S13360-3025CS  & 12.4               & 0.0        &  10.8                  & 21.64            & 1.00$^{\rm c}$ & 0.076                                         & 0.06         &   0.03       \\
S13360-1350CS  & 29.0               & 1.4        &   8.5                  &  6.09            & 1.13           & 0.022                                         & 0.01         & $<0.01$      \\
S13360-3050CS  & 28.5               & 0.0        &  11.4                  &  9.90            & 1.04           & 0.030                                         & 0.01         & $<0.01$      \\
\hline
\multicolumn{9}{l}{$^{\rm a}$ Value taken from the S12571-025C MPPC.}\\
\multicolumn{9}{l}{$^{\rm b}$ Value obtained by simulation.}\\
\multicolumn{9}{l}{$^{\rm c}$ Non-fitted value.}\\
\multicolumn{9}{l}{$^{\rm d}$ Value taken from the S12571-050C MPPC.}\\
\end{tabular}
\end{table}

\begin{figure}[t]
\centering
\includegraphics[width=.6\textwidth]{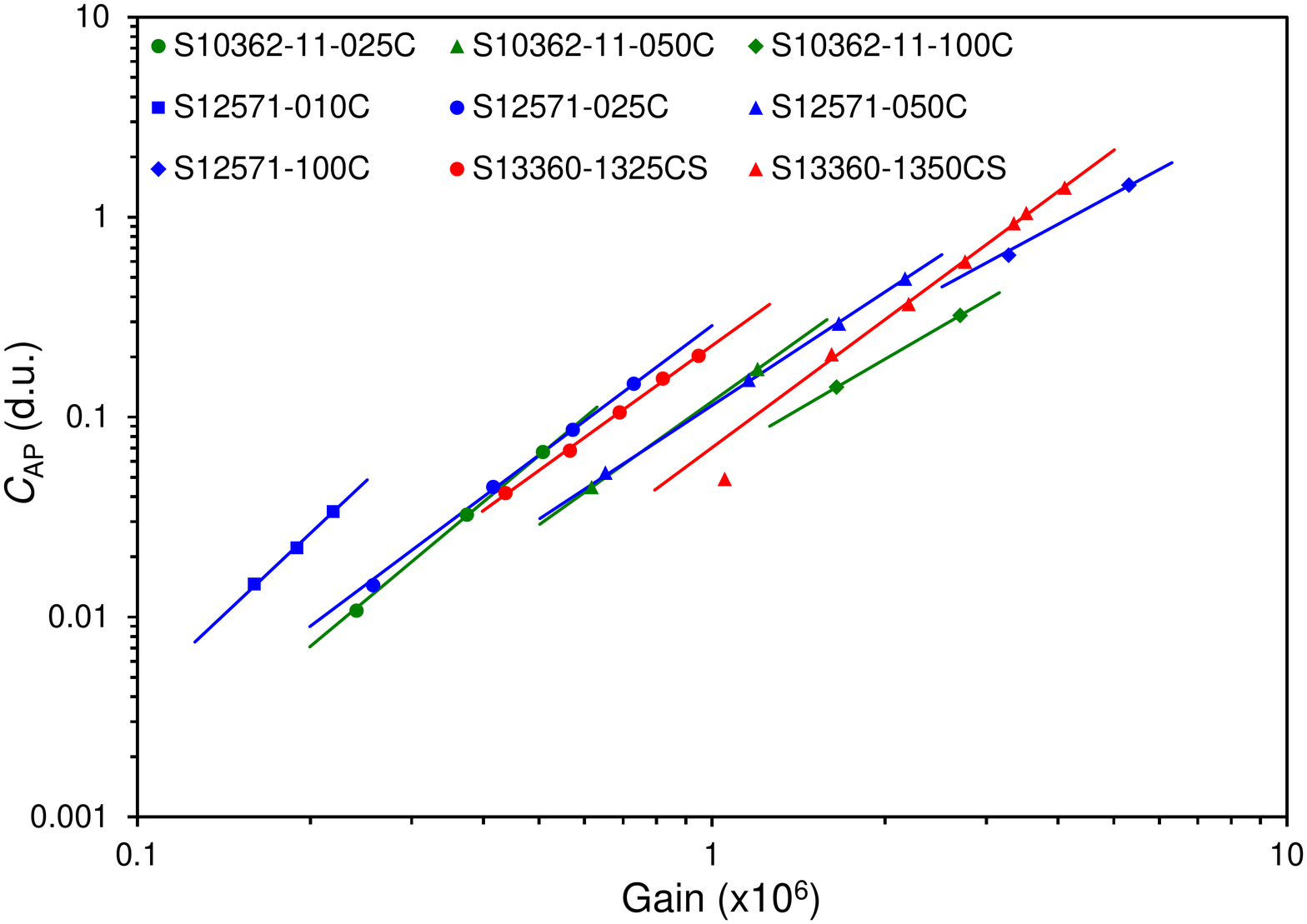}
\caption{Fitted $C_{\rm AP}$ values as a function of gain for the MPPCs with small active area.
The smaller the pixel size, the greater the $C_{\rm AP}$ value.
The lines represent power-law fits to data.}
\label{fig:C_AP}
\end{figure}

The $C_{\rm AP}$ and $C_{\rm CT}$ coefficients were found to have a power-law dependence with gain, but their ratio
keeps approximately constant. Since the two noise components are closely related to each other, we preferred to
characterize them by taking average values of $\tau_{\rm bulk}$ and $\frac{C_{\rm CT}}{C_{\rm AP}}$ for each MPPC
(table~\ref{tab:AP_results}) and refit $C_{\rm AP}$ as the only parameter depending on gain. The fitted $C_{\rm AP}$
values (dimensionless units) for most of the MPPCs are shown in figure~\ref{fig:C_AP}. MPPCs belonging to different
series but having the same pixel size have very close $C_{\rm AP}$ values at a given gain, despite the large
differences in their total amount of afterpulses. Similarly to the prompt crosstalk probability
(figure~\ref{fig:promptCT}), $C_{\rm AP}$ increases with decreasing pixel size, although the cause of this behavior is
not well understood yet. The $\frac{C_{\rm CT}}{C_{\rm AP}}$ ratio also shows a strong correlation with the pixel size,
which was studied by Monte Carlo simulation (section~\ref{sec:MC}). For some MPPCs where the components of afterpulsing
and delayed crosstalk could not be separated, we used either the $\frac{C_{\rm CT}}{C_{\rm AP}}$ value measured for
other MPPC with same pixel size or the value obtained by simulation.

The dependence of $C_{\rm AP}$ on gain or overvoltage is parameterized in a similar way to $\varepsilon$ in
eq.~(\ref{fit_CT}), with the difference that $C_{\rm AP}$ is not a probability and thus it may take values greater than
unity:

\begin{equation}
C_{\rm AP}=\left(\frac{X}{K_{\rm AP}}\right)^{(1+\beta)}\,.
\label{fit_C_AP}
\end{equation}
The best fits to data are shown in figure~\ref{fig:C_AP} and the $K_{\rm AP}$ and $\beta$ values are given in
table~\ref{tab:AP_results}. As expected, the $\beta$ indexes are similar to the $\alpha$ values obtained for
$\varepsilon$ (table~\ref{tab:fit_CT}). In some cases where the data were scarce, we set $\beta=1$.

To quantify the correlated noise of a MPPC, the mean number of secondary pulses of each type per primary avalanche can
be evaluated by integration of equations (\ref{f_AP_diff}) and (\ref{f_CT_diff}). The following integrals were
calculated for a reference gain of $1\cdot10^6$:

\begin{align}
N_{\rm AP}=& C_{\rm AP}\cdot\int_{3\,{\rm ns}}^{\infty}{\frac{1}{t}\cdot\exp\left(-\frac{t}{\tau_{\rm bulk}}\right)\cdot\mu(t)\cdot{\rm d}t}\nonumber\\
N_{\rm CT}=& C_{\rm CT}\cdot\int_{3\,{\rm ns}}^{\infty}{\frac{1}{\sqrt{t}}\cdot\exp\left(-\frac{t}{\tau_{\rm bulk}}\right)\cdot{\rm d}t}\,,
\label{N_sec}
\end{align}
where we assumed that the $f_{\rm AP}(t)$ and $f_{\rm CT}(t)$ functions can be extrapolated down to 3~ns. Note also that the
$k(t)$ factor in eq.~(\ref{f_AP_diff}) has been removed here to include all the afterpulses, regardless the
discrimination threshold. The results of these integrals are given in table~\ref{tab:fit_CT}. Both $N_{\rm AP}$ and
$N_{\rm CT}$ decrease by a factor of 5 or more from the old S10362-11 series to the S12571 series. A further reduction
of a factor of around 2 is observed in the S13360 series, but it is not as drastic as the reduction of prompt
crosstalk. Trenches seem to have no significant effect on delayed crosstalk.

We made a further test of hypothesis H2. The analysis was repeated for the S10362-11-050C and S13360-1350CS MPPCs but
selecting primary pulses with prompt crosstalk multiplicity 1, i.e., two simultaneous primary avalanches. As expected,
$\tau_{\rm bulk}$ maintained the same value and $N_{\rm AP}$ increased approximately by a factor of 2; more precisely
2.08 for the S10362-11-050C MPPC and 2.03 for the S13360-1350CS one. On the other hand, the increase in $N_{\rm CT}$
was only 35\% and 37\%, respectively. This could be justified by the fact that crosstalk between the two primary pixels
during their recovery gives rise to low-amplitude pulses that would be classified as afterpulses. A generalization of
our model to be applied to an arbitrary distribution of primary avalanches (e.g., triggered by a light pulse) should
include cascades of correlated pulses and the recovery of every fired pixel, not only the primary ones. Implementing
these features would also need to model what neighboring pixels are subject to delayed crosstalk for each fired pixel,
as done for the prompt crosstalk.

\section{Monte Carlo simulation}\label{sec:MC}

We developed a Monte Carlo simulation of the prompt crosstalk, afterpulsing and delayed crosstalk for MPPCs. The goal
of this simulation is twofold. In the first place, it has provided us with the information needed to parameterize the
time distributions $f_{\rm AP}(t)$ and $f_{\rm CT}(t)$ for modeling the afterpulsing and crosstalk due to minority
carriers diffusing in the silicon substrate (section~\ref{ssec:procedure}). In the second place, it allowed us to study
some aspects of the different components of crosstalk and afterpulsing, as their dependence on geometrical and physical
parameters and the impact of trenches.

The simulation includes the emission of secondary photons from a primary avalanche, their absorption in the different
parts of the device and the diffusion of photon-generated minority carriers in the substrate. Both photons and carriers
can trigger new avalanches, but the photons emitted by these secondary avalanches are not simulated, i.e., cascades are
ignored.

\subsection{Input physical parameters}\label{ssec:MC_parameters}

The wavelength spectrum of secondary photons was taken from~\cite{Mirzoyan}, which is reproduced in
figure~\ref{fig:wavelenght_spectra} in relative units. For simplicity, the distribution was sampled in 60 discrete
wavelength values between 450 and 1620~nm. This spectrum was measured for a MPPC of the S10362-11 series, although it
is expected to be valid for any SiPM. The absorption length of silicon at room temperature as a function of wavelength
was taken from~\cite{Green}.

The diffusion constant of minority carriers (holes) in the n-type substrate was varied within a large range from 0.1 to
1~$\mu{\rm m}^2$/ns, but we generally used a reference value of 0.4~$\mu{\rm m}^2$/ns corresponding to a doping density
equal or higher than $5\cdot10^{-18}$~cm$^{-3}$~\cite{Wang}.

\begin{figure}[t]
\centering
\includegraphics[width=.45\textwidth]{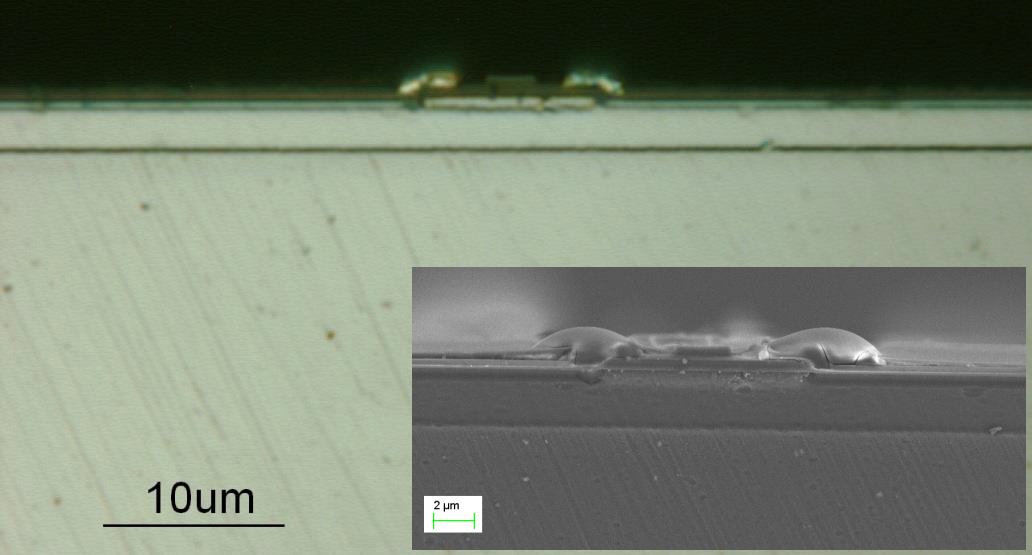}\,\,
\includegraphics[width=.45\textwidth]{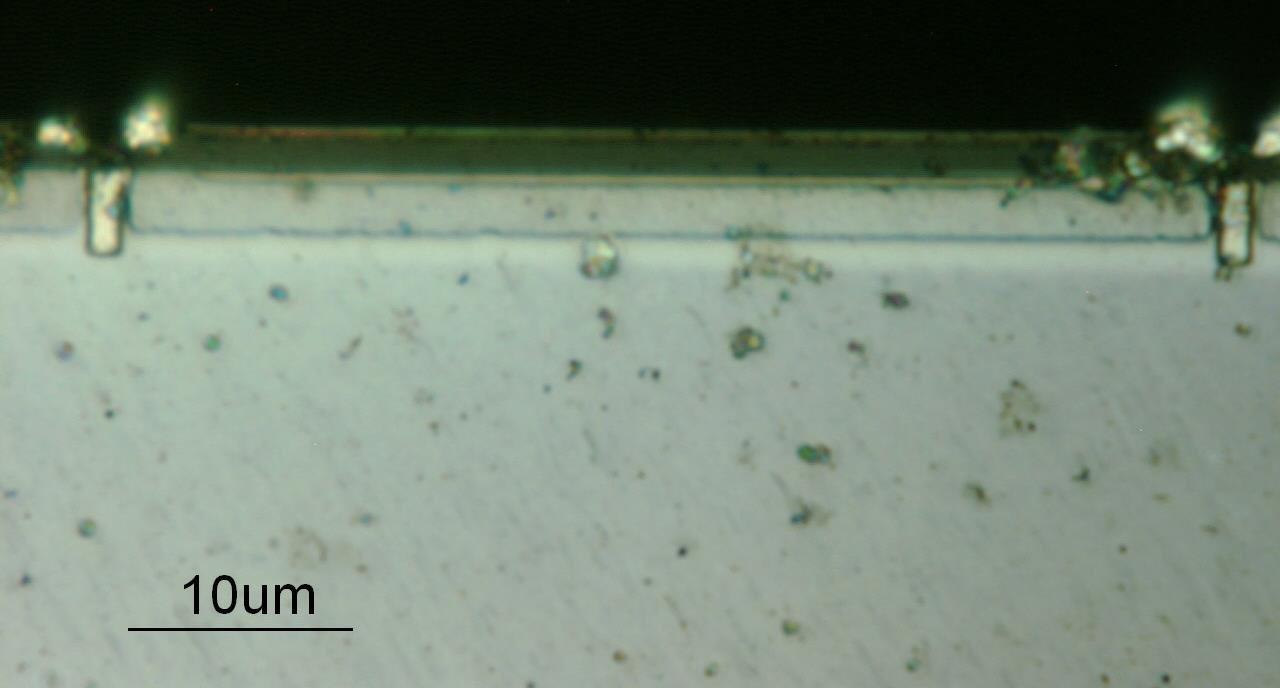}
\caption{Left-hand plot: optical image and scanning electron microscope image (inset) of a cross-section of a S12571-100C MPPC showing the boundary between two pixels.
Right-hand plot: optical image of a cross-section of a S13360-1350CS MPPC showing a pixel surrounded by optical trenches.
The p-n junction and the thin p++ electric contacts of the pixels can be distinguished.
The trenches of the S13360-1350CS MPPC extend down to the bottom edge of the avalanche layer to prevent direct optical crosstalk.}
\label{fig:profile}
\end{figure}

Some basic geometrical parameters of the MPPCs, as the pixel pitch $L_{\rm pix}$, the fill factor $F$ and the size of
the active area are provided by Hamamatsu. Besides, we carried out measurements on some MPPCs of the different series
to verify and complete this information. First, we decapsulated and recovered the silicon dice integrated in the
detector and measured the surface value of its geometrical parameters (e.g., die size, pixel size and separation
between adjacent pixels) using an optical microscope that achieves a magnification value of 1500. Then, by means of a
grinding and lapping metallographic process, we made several cross-sections in different areas of interest of the
analyzed sample, which were observed through both and optical microscope and a scanning electron microscope (SEM). This
allowed us to determine certain technical parameters to be implemented in our Monte Carlo code: thickness of the
substrate, detection zone depth and pattern, width and depth of the integrated wells, etc. Examples of these
cross-sections for the S12571-100C and S13360-1350CS MPPCs are in figure~\ref{fig:profile}.

All the MPPC series have a substrate with a thickness of around 630~$\mu$m and a similar layer structure of pixels. The
p-type layer has an estimated thickness of 2~$\mu$m. Between it and the substrate, there is a thin avalanche layer with
thickness $T_{\rm av}\approx1\,\mu$m, which we guessed to be lightly doped n type. In each pixel, a p++ contact ($\sim
0.5\,\mu$m thickness) is on the top of the p-type layer. The coating and passivating surface layers have an estimated
total thickness of 1.5~$\mu$m. The S13360 series have in addition trenches between pixels. As can be seen in
figure~\ref{fig:profile}, the trenches extend from the top edge of the p layer down to the bottom edge of the avalanche
layer. The particular pixel geometry of each MPPC was implemented in the Monte Carlo code.

\subsection{Algorithm}\label{ssec:MC_algorithm}

The simulation assumes that the primary avalanche takes place in a central pixel of the array. A predefined number of
secondary photons for each discrete wavelength are generated in proportion to the above-mentioned relative spectrum.
The initial photon position is uniformly distributed in the volume of the pixel avalanche region, which is defined by a
rectangular box of dimensions $L_{\rm av}\times L_{\rm av}\times T_{\rm av}$. We assumed two possible values of the
lateral size: $L_{\rm av}=L_{\rm pix}$ and $L_{\rm av}=L_{\rm pix}\cdot\sqrt{F}$, where $F$ is the fill factor. The
emission direction of the photon is isotropically distributed and its total distance traveled is generated with a
negative exponential distribution with parameter $\frac{1}{\lambda}$, where $\lambda$ is the absorption length of
silicon at the given wavelength. All the layers of the device, including the coating layer, are assumed to have the
same absorption length.

If a photon is absorbed in the active region of a pixel different from the primary one, it triggers a prompt-crosstalk
avalanche there. The active region is defined by a rectangular box with the same lateral size as the avalanche region
$L_{\rm av}$ and with a thickness comprising both the avalanche layer and the p layer, but not the p++ contact. A 100\%
trigger probability is assumed, since the simulation is only intended to obtain relative results on the prompt
crosstalk probability. Specular reflections on the top and bottom surfaces of the device are also simulated. For
simplicity, both surfaces are assumed to be perfectly smooth and to have 100\% reflectivity, regardless the photon
incident angle and wavelength. The reflected photons are tagged so that their contribution to crosstalk can be
evaluated separately.

For the S13360 series, if the photon trajectory intersects a trench, the photon is dumped and the simulation continues
to the next one. This means that direct crosstalk is not allowed taking into account the depth of the trenches
(right-hand plot of figure~\ref{fig:profile}). On the other hand, we considered no obstacle for transmission of photons
through the coating layer above a trench. Therefore, the trenches do not eliminate completely the crosstalk due to
photons reflected on the top-surface of the device, as will be shown below.

If a photon is absorbed in the substrate, it produces an electron-hole pair. The diffusion of the minority carrier,
i.e., the hole, is simulated as a random walk process with an step of 0.1~$\mu$m, but instead of forcing it to move in
a discrete grid, the direction of movement in each step is randomly generated with a uniform distribution. The elapsed
time is calculated as the number of steps times $\frac{(0.1\,\mu{\rm m})^2}{6\cdot D_{\rm h}}$, where $D_{\rm h}$ is
the diffusion constant of holes in the silicon substrate. The hole keeps diffusing until it reaches a border of the
substrate or the total elapsed time is longer than 300~ns, but recombination is ignored (i.e., $\tau_{\rm
bulk}\rightarrow\infty$). If the hole reaches the active region of a pixel, it produces instantaneously a secondary
avalanche and the event is classified as afterpulsing or crosstalk, depending on whether the pixel is the primary one
or not. Crosstalk events with delay time shorter than 3~ns are classified as prompt crosstalk. The trigger efficiency
of holes is also assumed to be 100\% and the recovery of the primary pixel is ignored. Generation and transport of
carriers in other parts of the device are not included either.

\subsection{Components of correlated noise}\label{ssec:MC_components}

\begin{figure}[t]
\centering
\includegraphics[width=.6\textwidth]{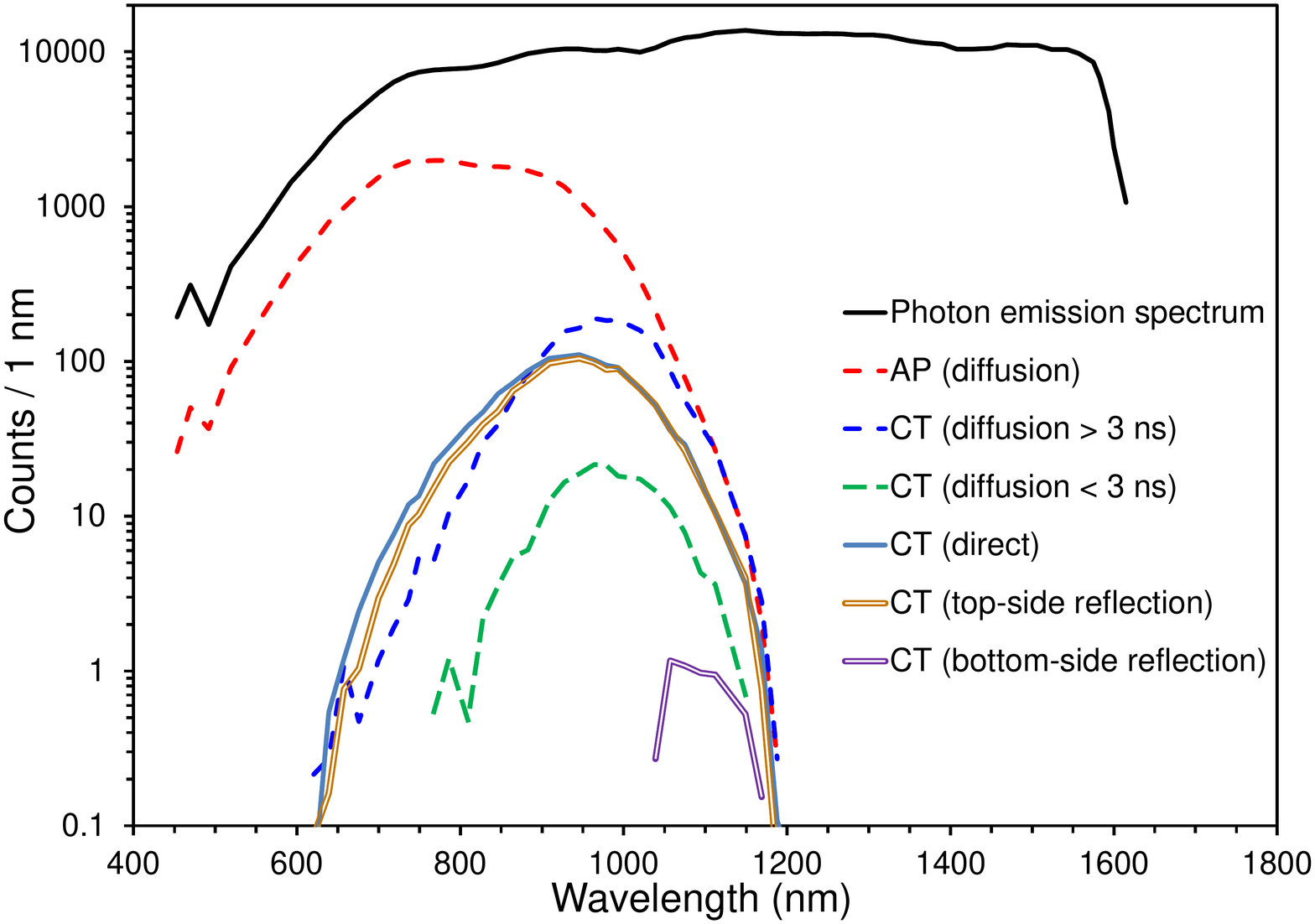}
\caption{Photon emission spectrum~\protect\cite{Mirzoyan} used in the simulation and the wavelength distributions of the different
components of crosstalk and afterpulsing obtained for the S12571-050C MPPC, assuming $L_{\rm av}=39.4\,\mu$m.
Results are for a statistics of $10^7$ simulated photons.}
\label{fig:wavelenght_spectra}
\end{figure}

In figure~\ref{fig:wavelenght_spectra}, we show the wavelength distributions of the simulated photons causing the
different components of afterpulsing and crosstalk for the S12571-050C MPPC, which has a pixel pitch of 50~$\mu$m. The
results were obtained assuming $D_{\rm h}=0.4\,\mu{\rm m}^2$/ns and $L_{\rm av}=39.4\,\mu$m (i.e., $F=62\%$). The
wavelength distributions for all the crosstalk components are within the range 600~--~1200~nm. This is determined by
the absorption length of silicon, which varies from a few $\mu$m at a wavelength of 600~nm to around 100~$\mu$m at
1000~nm and thereafter rises steeply~\cite{Green}. Photons with wavelength larger than 1200~nm are very unlikely to be
absorbed within the device, and those with wavelength shorter than 600~nm do not scape the primary pixel or are
absorbed very close to it, having a negligible probability to cause crosstalk. On the other hand, if assuming $L_{\rm
av}=50\,\mu$m, i.e., no ``dead zone'' between pixels, the number of crosstalk events increases approximately by a
factor of 4 and its wavelength distribution extends to shorter wavelengths. A similar but more moderate effect is
observed when decreasing the pixel size. Apart from that, the wavelength range of photons contributing to crosstalk
barely changes when varying other input parameters of the simulation. For instance, we obtained similar wavelength
distributions for the crosstalk components when assuming that the photon emission spectrum is that of a black body of
temperature 2000~K or 4500~K, as was also observed in~\cite{Otte}.

The prompt crosstalk due to top-side reflected photons is almost as high as the direct crosstalk, whereas the
contribution of bottom-side reflected photons is negligible. Obviously, photon reflections are highly overestimated in
the simulation, as we assumed 100\% reflectivity. However, it is worth considering the top-side reflections in the
design of SiPMs, e.g., the metal circuitry and the trenches. In fact, for the simplified geometry implemented in the
simulation for the S13360 series, we observed that the trenches reduce the crosstalk component due to top-side
reflected photons approximately by a factor of 4, but this would be completely eliminated if the barriers extend up to
the top surface of the device.

The simulation also overestimates the hole-induced afterpulsing and crosstalk, because it assumes an infinite bulk
lifetime and 100\% trigger probability for holes. This also distorts the shape of the wavelength distributions for
these components, because the photon wavelength is correlated with both the depth at which the photon is absorbed and
the time needed by the hole to reach a pixel. Most of the simulated holes causing afterpulsing and crosstalk are
generated at a depth less than 10~$\mu$m with respect to the bottom edge of the avalanche layer, although their depth
distribution extends to several tens of $\mu$m. This distribution becomes flatter if assuming a diffusion constant of
$1\,\mu{\rm m}^2$/ns instead of $0.4\,\mu{\rm m}^2$/ns. The holes contributing to prompt crosstalk (i.e., delay time
shorter than 3~ns) are those generated in a thin layer of 3~$\mu$m beneath the active region of pixels different to the
primary one. We found that the trenches of the S13360 series reduce the hole-induced crosstalk only by a few percent,
with the reduction in the prompt component being smaller than 10\%. Deeper trenches may reduce the crosstalk to a much
greater extent.

Unlike crosstalk, the wavelength distribution for afterpulsing in figure~\ref{fig:wavelenght_spectra} decreases with
decreasing wavelength approximately as the same rate as the photon emission spectrum does, because short-wavelength
photons produce holes very close to the primary pixel. However, this distribution was obtained ignoring the recovery of
the primary pixel. Including this would preferentially reduce the contribution of short-wavelength photons. On the
other hand, the pixel size and the $L_{\rm av}$ parameter were found to have a minor influence on the total number of
afterpulses per primary avalanche. The impact of trenches is also very small.

\subsection{Spatial distribution of crosstalk}\label{ssec:MC_CT}

\begin{figure}[t]
\centering
\includegraphics[width=.8\textwidth]{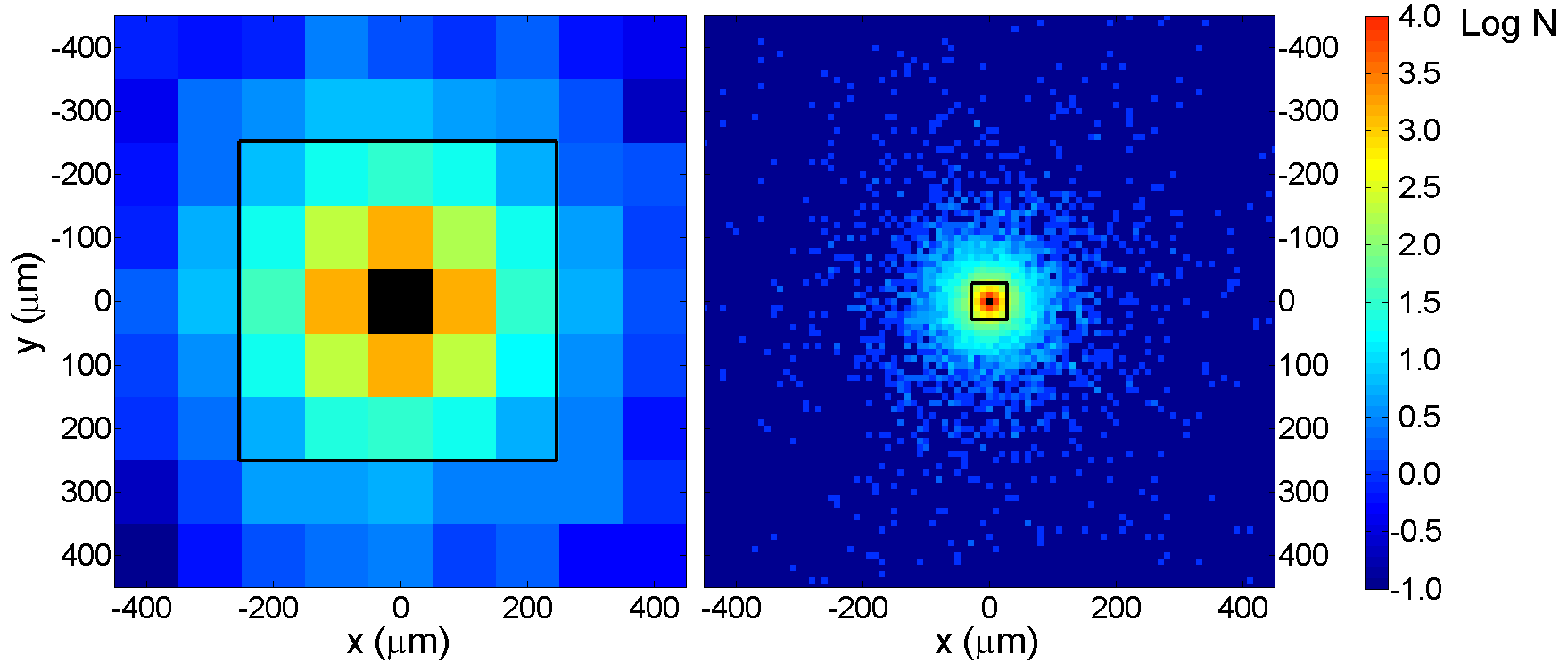}
\caption{Distribution of fired pixels via direct crosstalk for the S12571-100C (left) and S12571-010C (right) MPPCs.
The simulation assumes that the primary avalanche takes place in the central pixel and does not include crosstalk cascades.
The highlighted central square in each plot encloses 25 pixels.
Results were obtained for a statistics of $10^7$ simulated secondary photons.}
\label{fig:maps_12571}
\end{figure}

We obtained for each crosstalk component the 2D histogram of fired pixels for a primary avalanche triggered in the
central pixel of the array. In figure~\ref{fig:maps_12571}, the spatial distribution of direct crosstalk obtained for
the 100 and 10~$\mu$m pixel pitch MPPCs of the S12571 series are shown, where $L_{\rm av}$ was assumed to be 88.3 and
$5.74\,\mu$m, respectively. Most of the crosstalk avalanches are triggered in the 24 pixels surrounding the primary
pixel (i.e., the highlighted central square in each plot). In particular, the fraction of avalanches in this area is
92\% (82\%) for the S12571-100C (-010C) MPPC, with 84\% (67\%) of total being in the 8 nearest pixels and 71\% (50\%)
in the 4 nearest ones. Assuming $L_{\rm av}=L_{\rm pix}$ leads to the fractions 97\% (89\%), 93\% (78\%) and 85\%
(62\%), in the same order. As observed with experimental data in section~\ref{ssec:CT_multiplicity}, the larger the
pixel size, the lower the number of neighboring pixels subject to direct crosstalk. Similar results were obtained for
the prompt crosstalk component due to top-side reflected photons. On the other hand, for the S13360 MPPCs, the trenches
eliminate the direct crosstalk and only allow photons with small incident angle to be reflected on the top surface. As
a consequence, the totality of avalanches due to top-side reflected photons take place in the 8 nearest pixels.

The spatial distribution for hole-induced crosstalk is more spread out. For instance, for the S12571-010C MPPC, the
fraction of avalanches with delay time shorter than 300~ns in the 24 nearest neighbors of the primary pixel is only
63\%, and the fraction increases to 69\% if the delay time is limited to 3~ns, i.e., the prompt component.

Using these simulation results, we calculated the prompt crosstalk multiplicity distribution to be compared directly
with the experimental data described in section~\ref{ssec:CT_multiplicity}. This was carried out by means of a Monte
Carlo algorithm that generates a large number of random cascades of prompt crosstalk, each one initiated by firing a
random pixel of the MPPC. During a cascade, the probability that a fired pixel induces prompt crosstalk in other pixel
(not fired yet) is proportional to the counts registered in the corresponding relative position in the above 2D
histogram obtained by simulation. The proportionality constant is determined so that the sum of the crosstalk
probabilities of all the pixels gives $\varepsilon$. Actually, this sum depends on the position of the primary pixel in
the array. Thus, we calculated it for all the possible primary pixel positions and then made an average. The
S12571-050C MPPC was chosen for this analysis, since we measured an amplitude spectrum with many peaks and good
resolution for it (figure~\ref{fig:spectrum_height_12571}). Only the 2D histogram of simulated direct crosstalk events
was used in the calculation, because the top-side reflected photons lead to a very similar spatial distribution, as
explained above, and the contribution of holes to prompt crosstalk is supposed to be small for this MPPC. For
simplicity, this 2D histogram was averaged over all the pixels outside the square enclosing the primary pixel and its
24 nearest neighbors, that is, we assumed that a fired pixel has the same probability to induce prompt crosstalk in any
pixel that is not within this area. This approximation can be done because only 10\% of the simulated prompt crosstalk
avalanches take place outside this area of pixels for the S12571-050C MPPC. We assumed $L_{\rm av}=39.4\,\mu$m (i.e.,
$F=62\%$) in the simulation.

\begin{figure}[t]
\centering
\includegraphics[width=.6\textwidth]{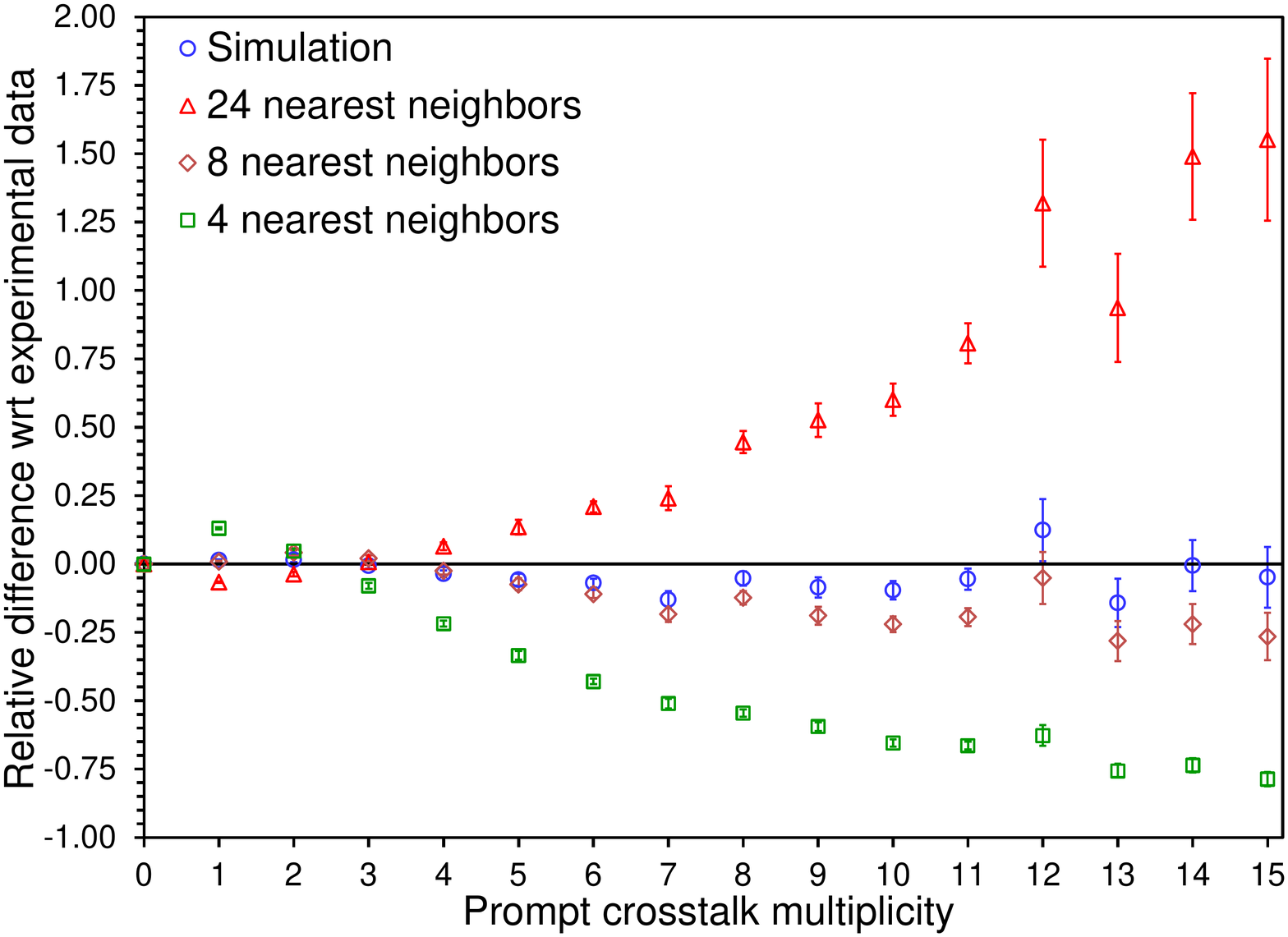}
\caption{Comparison of the prompt crosstalk multiplicity distribution calculated using simulation results of direct crosstalk
for the S12571-050C MPPC with the experimental distribution obtained from the pulse amplitude spectrum of figure~\protect\ref{fig:spectrum_height_12571}.
The distributions calculated for three simple geometrical assumptions
(i.e., the prompt crosstalk can only take place in the 4, 8 or 24 nearest neighbors of a fired pixel, with the same probability for all these neighbors)
are also included in the comparison.}
\label{fig:simulation_CT_12571}
\end{figure}

The relative difference of the calculated prompt crosstalk multiplicity with respect to the experimental one for the
S12571-050C MPPC is represented by blue circles in figure~\ref{fig:simulation_CT_12571}. A remarkable agreement is
observed, although the difference for some multiplicities is larger than the error bars, which include both the
experimental uncertainties and the statistical uncertainties of the numerical calculation. We also show the comparison
for three multiplicity distributions calculated in the same way, but instead of using the simulation results, assuming
that the prompt crosstalk can only take place respectively in the 4, 8 or 24 nearest neighbors of a fired pixel and
that all these neighbors have the same crosstalk probability. Unlike the analytical formulae used in
section~\ref{ssec:CT_multiplicity}, these numerical calculations allow the determination of the distribution for
arbitrarily high multiplicities and take into account that border pixels have a smaller number of neighbors subject to
prompt crosstalk.

\subsection{Time distribution of afterpulsing and delayed crosstalk}\label{ssec:MC_aft}

\begin{figure}[t]
\centering
\includegraphics[width=.6\textwidth]{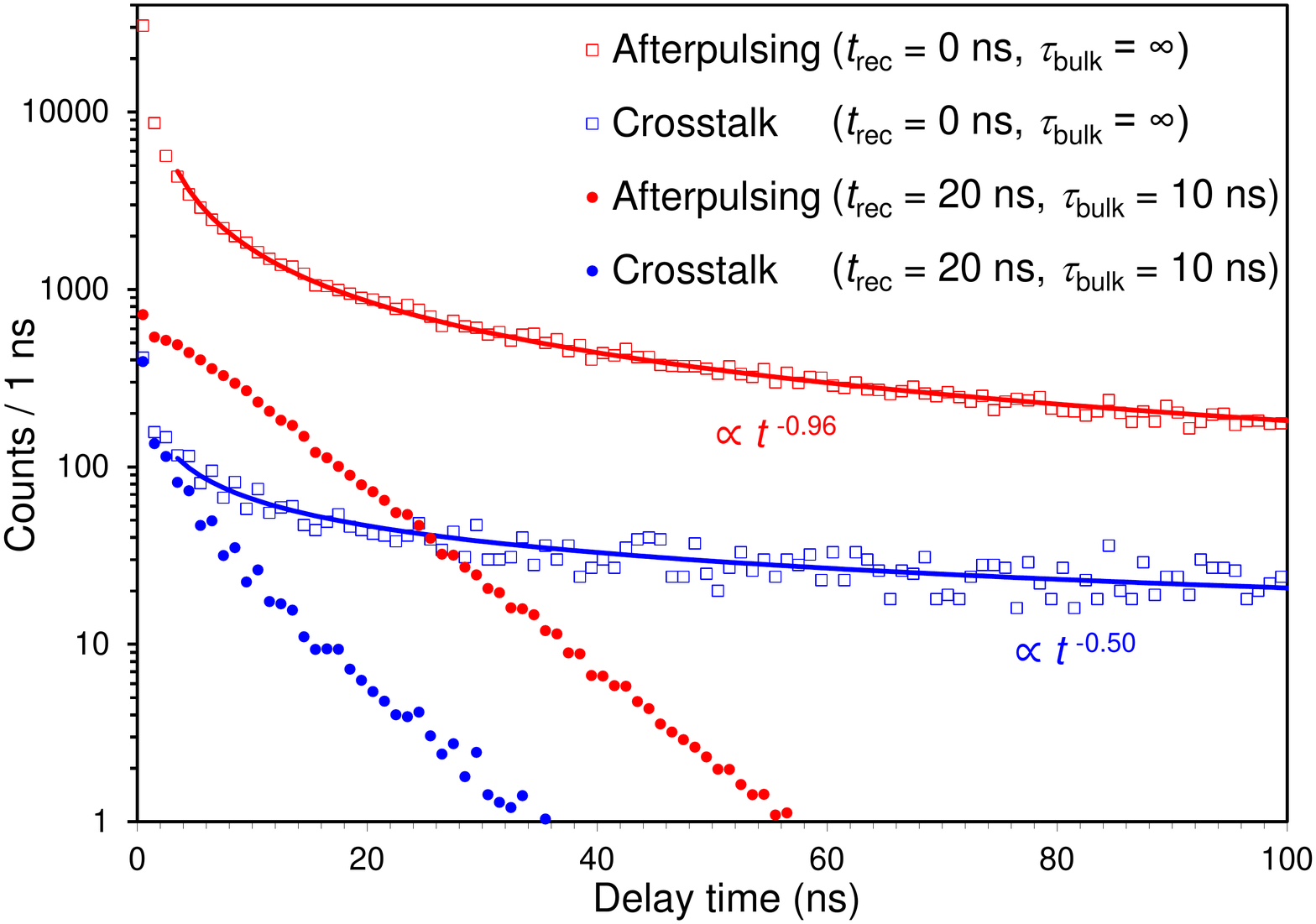}
\caption{Diffusion time distribution of simulated holes that cause afterpulsing (red squares) and crosstalk (blue squares) for the S12571-050C MPPC.
Solid lines represent power law fits.
Filled circles represent the time distribution after including the effects of the primary pixel recovery and hole recombination in the substrate,
assuming $\tau_{\rm rec}=20$~ns and $\tau_{\rm bulk}=10$~ns.}
\label{fig:MC_diffusion}
\end{figure}

Our simulation also provides the diffusion time distribution of simulated holes that cause afterpulsing and crosstalk.
Results for the S12571-050C MPPC, assuming $L_{\rm av}=39.4\,\mu$m and $D_{\rm h}=0.4\,\mu{\rm m}^2$/ns, are shown in
figure~\ref{fig:MC_diffusion}. A power law fit to data between 3 and 100~ns is shown for both components, with the
fitted exponent being $a=-0.96$ for afterpulsing and $b=-0.50$ for delayed crosstalk. Similar $a$ and $b$ values are
obtained for other simulation input data, where $a$ ranges from -0.83 to -1.25 and $b$ ranges from -0.24 to -0.56. A
slight decrease in $|a|$ as increasing pixel size or decreasing diffusion constant is observed. The dependence of $b$
on these parameters is not clear, but trenches make the time distribution for crosstalk flatter, with $b>-0.30$,
because they intercept some laterally emitted photons that would generate holes close to the avalanche layer (i.e.,
short diffusion time). In general, power law functions with exponents $a=-1.0$ and $b=-0.5$ fit well to the simulation
data and also allow us to describe the experimental delay time distribution of secondary pulses, as described in
section~\ref{sec:aft}.

As mentioned above, neither the recovery of the primary pixel or the recombination of holes in the substrate are
simulated. In figure~\ref{fig:MC_diffusion}, we also show the time distributions of afterpulsing and crosstalk after
including these effects as described in section~\ref{ssec:procedure}, where $\tau_{\rm rec}=20$~ns and $\tau_{\rm
bulk}=10$~ns are assumed.

We calculated the relative contributions of afterpulsing and delayed crosstalk as a function of the pixel size. To do
that, we refitted the simulation data assuming $a=-1.0$ and $b=-0.5$. The following $\frac{C_{\rm CT}}{C_{\rm AP}}$
ratios were obtained for the MPPCs of the S12571 series: 0.14 for $L_{\rm pix}=10\,\mu$m, 0.030 for $L_{\rm
pix}=25\,\mu$m, 0.013 for $L_{\rm pix}=50\,\mu$m and 0.006 for $L_{\rm pix}=100\,\mu$m, which are in general agreement
with the ratios obtained experimentally in section~\ref{ssec:AP_results}. The decrease in $\frac{C_{\rm CT}}{C_{\rm
AP}}$ with increasing pixel size is due to the fact that short-wavelength photons contribute differently to the
afterpulsing and crosstalk components, as explained in section~\ref{ssec:MC_components}. The results are only slightly
dependent on the input parameters of the simulation, e.g., $L_{\rm av}$ and $D_{\rm h}$.

\section{Conclusions}\label{sec:conclusions}

The correlated noise in several Hamamatsu MPPCs was analyzed in depth. We considered the noise contribution of the
secondary photons emitted by avalanches as well as of the photon-generated minority carriers diffusing in the silicon
substrate of the device. Our experimental method was based on a waveform analysis that is able to identify pulses
with time differences as low as 3~ns. This determined our empirical criterium to differentiate between prompt and
delayed crosstalk. In addition, we used the fact that the afterpulse amplitude is lowered during the primary pixel
recovering to characterize the afterpulsing separately from the delayed crosstalk.

The prompt crosstalk was characterized experimentally through the pulse amplitude spectrum at dark conditions. In
particular, the gain dependence of the prompt crosstalk probability was measured and parameterized. The new S13360
series of MPPCs, which incorporate isolating trenches between pixels, were found to have a prompt crosstalk probability
more than a factor of 30 lower than the former MPPC series. However, it is still non-negligible, likely due to both
secondary photons reflected on the top surface of the MPPC and minority carriers diffusing in the substrate. Using the
information provided by the prompt crosstalk multiplicity distribution and analytical models proposed in a previous
work~\cite{Gallego}, we deduced that prompt crosstalk mostly takes place in a small area of pixels ($\sim 8$) around
the primary one.

A novel procedure to characterize the afterpulsing and delayed crosstalk was presented in this work. It uses both the
amplitude and delay time distributions of secondary pulses. The primary pixel recovery, the pulse detection threshold
and the prompt crosstalk in secondary avalanches were carefully taken into account. In addition, the time distributions
of both afterpulses and delayed-crosstalk pulses due to diffused minority carriers were modeled using results from a
Monte Carlo simulation. A significant component of delayed crosstalk was observed in all the MPPCs that were
characterized. Also, our results indicate that the dominant contribution of afterpulsing is due to diffused minority
carriers, not to the release of avalanche carriers trapped in deep levels of the depleted layer, as commonly believed.
In fact, we attributed the drastic afterpulse reduction observed in the new MPPC series to an increase in the doping
density of the n-type silicon substrate from 3~--~6~$\cdot10^{18}\,{\rm cm}^{-3}$ to $2\cdot10^{19}\,{\rm cm}^{-3}$,
involving a notably decrease of the minority carrier lifetime from a few hundreds of ns to around 10~ns. The recovery
time was also shown to have an important role in the afterpulse reduction.

We developed a Monte Carlo simulation of the physical processes causing afterpulsing and crosstalk. It includes the
emission and transport of secondary photons as well as the diffusion of photon-generated minority carriers (holes) in
the substrate. We measured the geometrical parameters of the various MPPCs to be implemented in the code. The
simulation allowed us to analyze in detail the different components of afterpulsing and crosstalk, e.g., the wavelength
distribution of secondary photons related to each noise component, the spatial distribution of crosstalk events, and
the diffusion time distribution of minority carriers causing afterpulsing and delayed crosstalk. The simulation results
are in general agreement with experimental data and support our hypotheses for their interpretation. We also concluded
that, although the trench structure used in the S13360 MPPCs proved to reduce drastically the prompt crosstalk, longer
trenches extending from the top surface of the device down to several $\mu$m into the substrate would reduce both
prompt and delayed crosstalk to a much greater extent.

\acknowledgments

This work was supported by the Spanish Ministry of Economy and Competitiveness through contracts FPA2012-39489-C04-02,
FPA2013-48387-C6-2-P and FPA2014-55295-C3-2-R. The authors would like to thank to Hamamatsu Photonics K.K.\ for its
technical support, Roger Dur\'{a} for the preparation of MPPC samples for the measurement of their geometrical
parameters, and Fernando Arqueros for the careful revision of this manuscript.

\end{document}